\documentclass[%
 reprint,
nofootinbib,
 amsmath,amssymb,
 aps,showkeys,
]{revtex4-1}

\usepackage{graphicx}
\usepackage{dcolumn}
\usepackage{bm}
\usepackage{textpos}
\usepackage{hyperref}
\usepackage{amsmath}
\usepackage{breqn}
\usepackage{mhchem}
\usepackage{wasysym}
\setcitestyle{authoryear,open={(},close={)}}
\renewcommand{\thetable}{\arabic{table}}

\makeatletter
\let\@fnsymbol\@arabic
\makeatother

\newcommand{\beginappendixa}{%
        \setcounter{table}{0}
        \renewcommand{\thetable}{A\arabic{table}}%
        \setcounter{figure}{0}
        \renewcommand{\thefigure}{A\arabic{figure}}%
        \setcounter{equation}{0}
        \renewcommand{\theequation}{A\arabic{equation}}%
     }
     
\newcommand{\beginappendixb}{%
        \setcounter{table}{0}
        \renewcommand{\thetable}{B\arabic{table}}%
        \setcounter{figure}{0}
        \renewcommand{\thefigure}{B\arabic{figure}}%
        \setcounter{equation}{0}
        \renewcommand{\theequation}{B\arabic{equation}}%
     }
     
\newcommand{\beginappendixc}{%
        \setcounter{table}{0}
        \renewcommand{\thetable}{C\arabic{table}}%
        \setcounter{figure}{0}
        \renewcommand{\thefigure}{C\arabic{figure}}%
        \setcounter{equation}{0}
        \renewcommand{\theequation}{C\arabic{equation}}%
     }
     
\newcommand{\beginappendixd}{%
        \setcounter{table}{0}
        \renewcommand{\thetable}{D\arabic{table}}%
        \setcounter{figure}{0}
        \renewcommand{\thefigure}{D\arabic{figure}}%
        \setcounter{equation}{0}
        \renewcommand{\theequation}{D\arabic{equation}}%
     }

\begin{document}

\widetext

\preprint{APS/123-QED}


\makeatletter
\providecommand\add@text{}
\newcommand\tagaddtext[1]{%
  \gdef\add@text{#1\gdef\add@text{}}}%
\renewcommand\tagform@[1]{%
  \maketag@@@{\llap{\add@text\quad}(\ignorespaces#1\unskip\@@italiccorr)}%
}
\makeatother

\newcommand{\beginsupplement}{%
        \setcounter{table}{0}
        \renewcommand{\thetable}{S\arabic{table}}%
        \setcounter{figure}{0}
        \renewcommand{\thefigure}{S\arabic{figure}}%
        \setcounter{equation}{0}
        \renewcommand{\theequation}{S\arabic{equation}}%
     }
     
\renewcommand{\thetable}{\arabic{table}}

\title{Meteoritic Abundances of Fatty Acids and Potential Reaction Pathways in Planetesimals}

\author{James C.-Y. Lai\textsuperscript{1}}
\author{Ben K. D. Pearce\textsuperscript{1,2}}
\author{Ralph E. Pudritz\textsuperscript{1,3}}
\author{Drake Lee\textsuperscript{1}}

\begin{flushright}
\text{{\it Accepted for publication in Icarus Sept. 24, 2018}}
\end{flushright}

\footnotetext[1]{Origins Institute and Department of Physics and Astronomy, McMaster University, ABB 241, 1280 Main St, Hamilton, ON, L8S 4M1, Canada}
\footnotetext[2]{Corresponding author: pearcbe@mcmaster.ca}
\footnotetext[3]{pudritz@mcmaster.ca}

\begin{abstract}
The origin of fatty acids on the prebiotic Earth is important as they likely formed the encapsulating membranes of the first protocells. Carbon-rich meteorites (i.e., carbonaceous chondrites) such as Murchison and Tagish Lake are well known to contain these molecules, and their delivery to the early planet by intense early meteorite bombardments constitutes a key prebiotic source. We collect the fatty acid abundances measured in various carbonaceous chondrites from the literature and analyze them for patterns and correlations. Fatty acids in meteorites include straight-chain and branched-chain monocarboxylic and dicarboxylic acids up to 12 carbons in length---fatty acids with at least 8 carbons are required to form vesicles, and modern cell membranes employ lipids with $\sim$12--20 carbons. To understand the origin of meteoritic fatty acids, we search the literature for abiotic fatty acid reaction pathways and create a candidate list of 11 reactions that could potentially produce these fatty acids in meteorite parent bodies. Straight-chain monocarboxylic acids (SCMA) are the dominant fatty acids in meteorites, followed by branched-chain monocarboxylic acids (BCMA). SCMA are most abundant in CM2 and Tagish Lake (ungrouped) meteorites, ranging on average from 10$^2$ ppb to 4$\times$10$^5$ ppb, and 10$^4$ ppb to 5$\times$10$^6$ ppb, respectively. In CM, CV, and Tagish Lake meteorites, SCMA abundances generally decrease with increasing carbon chain length. Conversely, SCMA abundances in CR meteorites peak at 5 and 6 carbons in length, and decrease on either side of this peak. This unique CR fatty acid distribution may hint at terrestrial contamination, or that certain fatty acid reactions mechanisms are active in different meteorite parent bodies (planetesimals). We identify Fischer-Tropsch-type synthesis as the most promising pathway for further analysis in the production of fatty acids in planetesimals.
\end{abstract}



\keywords{astrobiology $|$ fatty acids $|$ meteorites $|$ reaction pathways $|$ planetesimals}

\maketitle

\section{Introduction}

Primitive cells (protocells) are a key component in the emergence of life as they allow for the concentration and protection of organics from the external environment, and the evolution of genetic material \citep{Reference428,Reference87}. Protocells have two key components: an information polymer and an encapsulating membrane \citep{Reference410}. In the RNA world hypothesis, the first self-replicating information polymers to form on Earth were RNA \citep{1986Natur.319..618G,2013AsBio..13..391N,Reference425}, and the first encapsulating membranes were likely fatty acid vesicles \citep{Reference87,Reference410,Reference428,2002AsBio...2..371D}. 

As with the amino acids and nucleobases, the presence of fatty acids in meteorites suggests a planetesimal origin of these molecules. (Planetesimals are approximately 1--100 km-sized bodies originating from the protoplanetary disk.) Fatty acids are the most abundant water-soluble organic compounds in carbon-rich meteorites (i.e., carbonaceous chondrites) \citep{2005GeCoA..69.1073H}. When immersed in water, these molecules spontaneously self-assemble into bilayers, which in turn form vesicles \citep{2002AsBio...2..371D,Reference431,1989OLEB...19...21D}. 

While the majority of life today uses phospholipids in its cellular membranes, it is expected that the original molecules in membranes were smaller and less complex \citep{2002AsBio...2..371D}. Lipids in modern cell membranes generally have carbon chains 12--20 carbons long \citep{1989OLEB...19...21D}, whereas vesicles can be composed of a single fatty acid type as short as 8 carbons in length \citep{Reference429}. Meteorites contain fatty acids 2--12 carbons in length; therefore vesicles could indeed form directly out of meteorite-delivered fatty acids \citep{1999OLEB...29..187N,Reference413,Reference412}.

In previous works, we collected and compared the abundances of amino acids \citep{2014ApJ...783..140C} and nucleobases \citep{Reference46} within different types of carbonaceous chondrites. We then simulated the formation of these biomolecules within meteorite parent bodies and compared the resulting abundances with the meteoritic data. We determined the dominant reaction pathways for amino acids and nucleobases to be Strecker synthesis \citep{Reference44} and Fischer-Tropsch-type synthesis \citep{Reference106}, respectively.

In this paper we extend our analyses of organics within meteorites to fatty acids by first collecting from the published literature the abundances of these molecules across the various carbonaceous chondrite types. We focus our analysis on 4 types of aliphatic fatty acids: straight-chain monocarboxylic acids (SCMA), branched-chain monocarboxylic acids (BCMA), straight-chain dicarboxylic acids (SCDA), and branched-chain dicarboxylic acids (BCDA). However, because the majority of meteoritic fatty acid studies to date have focused on measuring SCMA, the majority of our data and analysis are on this fatty acid type. A few short-chain aromatic carboxylic acids are also present in carbonaceous chondrites \citep{2006MPS...41.1073M}. Although the permeation of these molecules through lipid bilayers has been studied \citep{Reference448}, such molecules are not thought to be incorporated into protocells. We analyze the fatty acid abundances in meteorites for patterns and search for correlations between fatty acids and other fatty acids, fatty acids and amino acids, and fatty acids and nucleobases.

We then perform a survey of the reaction routes that potentially formed these molecules within meteorite parent bodies. We focus on reactions that could occur in planetesimal interiors, and whose reactants are likely to have been present within such environments.


In section~\ref{sec2}, we summarize the quantitative analyses of fatty acids in meteorites, and display the meteoritic abundances of fatty acids in search of patterns. In this section, we also look for correlations between abundances of straight-chain and branch-chain fatty acids, and compare abundances of fatty acids with those of glycine and guanine in the same meteorites. Then, in section~\ref{sec3}, we list all well-known abiotic reaction mechanisms which form the types of fatty acids found in meteorites. We then reduce these mechanisms to a candidate list of reaction pathways which may have produced fatty acids in meteorite parent bodies. Next, in section~\ref{sec4}, we suggest the dominant fatty acid synthesis pathways in these bodies based on the reactant abundances available in planetesimal interiors. In section~\ref{sec5}, we discuss the distinct fatty acid distributions in CR and CM meteorites (see Table~\ref{glossary} for glossary of terms). Finally, in section \ref{sec7}, we summarize the main findings in this paper and state the future direction of our work.

\begin{table}
	\caption{Glossary of terms.\label{glossary}}
	\centering
	\begin{tabular}{cl}
		Term & Definition \\
		\hline \\[-2.5mm]
		SCMA & straight-chain monocarboxylic acid\\
		BCMA & branched-chain monocarboxylic acid\\
		SCDA & straight-chain dicarboxylic acid\\
		BCDA & branched-chain dicarboxylic acid\\
		CM & Mighei-type carbonaceous chondrite\\
		CR & Renazzo-type carbonaceous chondrite\\
		C2 & Ungrouped carbonaceous chondrite of petrologic type 2\\
		\tableline
		
	\end{tabular}
\end{table}

\section{Meteoritic Fatty Acids}\label{sec2}

\begin{figure}
	\includegraphics[width=5cm]{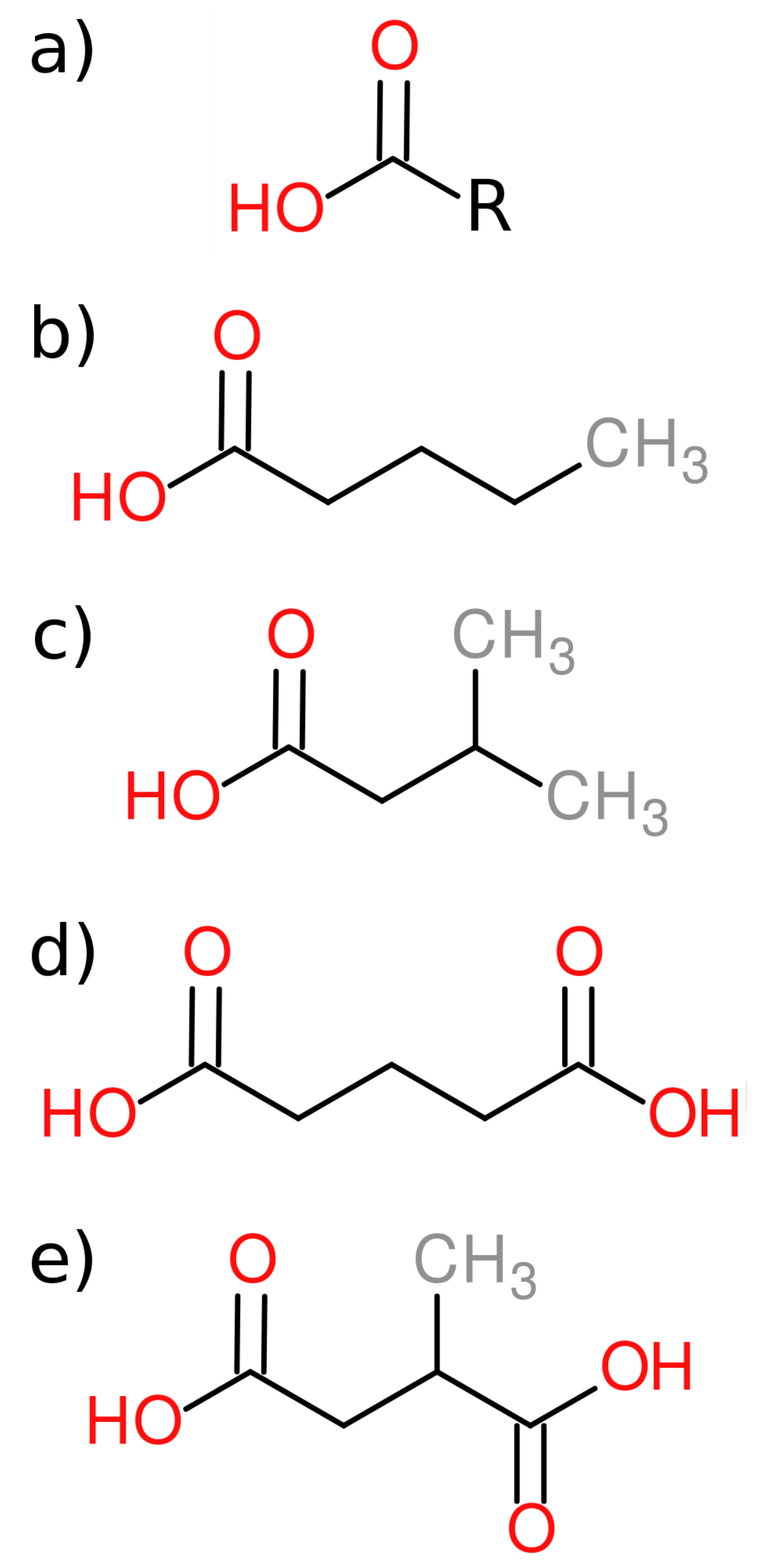}
	\caption{{\bf a)} The general form of a carboxylic acid, and examples of {\bf b)} a straight-chain monocarboxylic acid (SCMA), {\bf c)} a branched-chain monocarboxylic acid (BCMA), {\bf d)} a straight-chain dicarboxylic acid (SCDA), and {\bf e)} a branched-chain dicarboxylic acid (BCDA). Carboxylic acids are amphiphiles, consisting of a straight or branched hydrophobic carbon chain and one or more hydrophilic carboxyl groups. Molecules b)--e) are pentanoic acid, 3-methylbutyric acid, pentanedioic acid, and methylbutanedioic acid, respectively. \label{Fig1}}
\end{figure}

Fatty acids are generally hydrocarbon chains capped by a carboxyl group (C(=O)OH). Diotic fatty acids (dicarboxylic acids) are capped by two carboxyl groups. The majority of fatty acids found in carbonaceous chondrites are monocarboxylic acids; however there is also limited data on dicarboxylic acids. The general form of a carboxylic acid is represented in Figure~\ref{Fig1}a, followed by examples of a straight-chain monocarboxylic acid (SCMA), a branched-chain monocarboxylic acid (BCMA), a straight-chain dicarboxylic acid (SCDA), and a branched-chain dicarboxylic acid (BCDA) in Figure~\ref{Fig1}b--e. R represents the branched or unbranched hydrocarbon chain potentially capped by a carboxyl group. For example, in Figure~\ref{Fig1}b, R represents a hydrocarbon chain of length four. In branched-chain fatty acids, hydrocarbon chains can branch off of the main hydrocarbon tree at one or more locations. Varying the length and the branching of the R chain varies the identity of the carboxylic acid. 


\subsection{Meteoritic Fatty Acid Data}

We search the literature for all analyses of fatty acids in meteorites to date and compile the reported abundances. Fatty acids have been detected and quantified in CR, CM, CV and C (ungrouped) carbonaceous chondrites, spanning petrologic types 1--3. Meteorites are classified into petrologic types based on their level of hydrothermal alteration, the significance being that aqueous-based chemistry could have occurred in their parent bodies. For a summary on meteorite classification, see Appendix A. We present a summary of the fatty acid meteoritic data in Table~\ref{tbl-data} in Appendix B.

Analytical and extraction techniques differ from study to study, which may vary the losses and accuracy of the measured fatty acid abundances. Furthermore, meteoritic samples are susceptible to low levels of contamination during sample storage \citep{2011GeCoA..75.2309A}. In one sample case (Tagish Lake 11v), small concentrations of contaminant have been measured \citep{2014MPS...49..526H}. However, deuterium enrichments, the lack of dominant hexadecanoic acid (SM16) and octadecanoic acid (SM18) abundances, and the low levels of amino acid contamination in the analyzed meteorite samples suggest that the abundances of fatty acids in these studies are largely extraterrestrial in origin. Moreover, the decreasing fatty acid abundances for increasing carbon-chain length in the C2 (Tagish), CM, and CV meteorites is indicative of an extraterrestrial origin \citep{1979Natur.282..396L,Reference442}. For the CR meteorites, the peak in fatty acid abundances at 5--6 carbons may suggest some level of terrestrial contamination. Increased ambiguity also exists for the CR samples due to their lack of deuterium data. For more details on the experimental methods used in the fatty acid analyses, and a discussion on contamination, see Appendices B and C, respectively.

\subsection{Fatty Acid Abundances in Meteorites}

\begin{figure*}
	\includegraphics[width=0.8\linewidth]{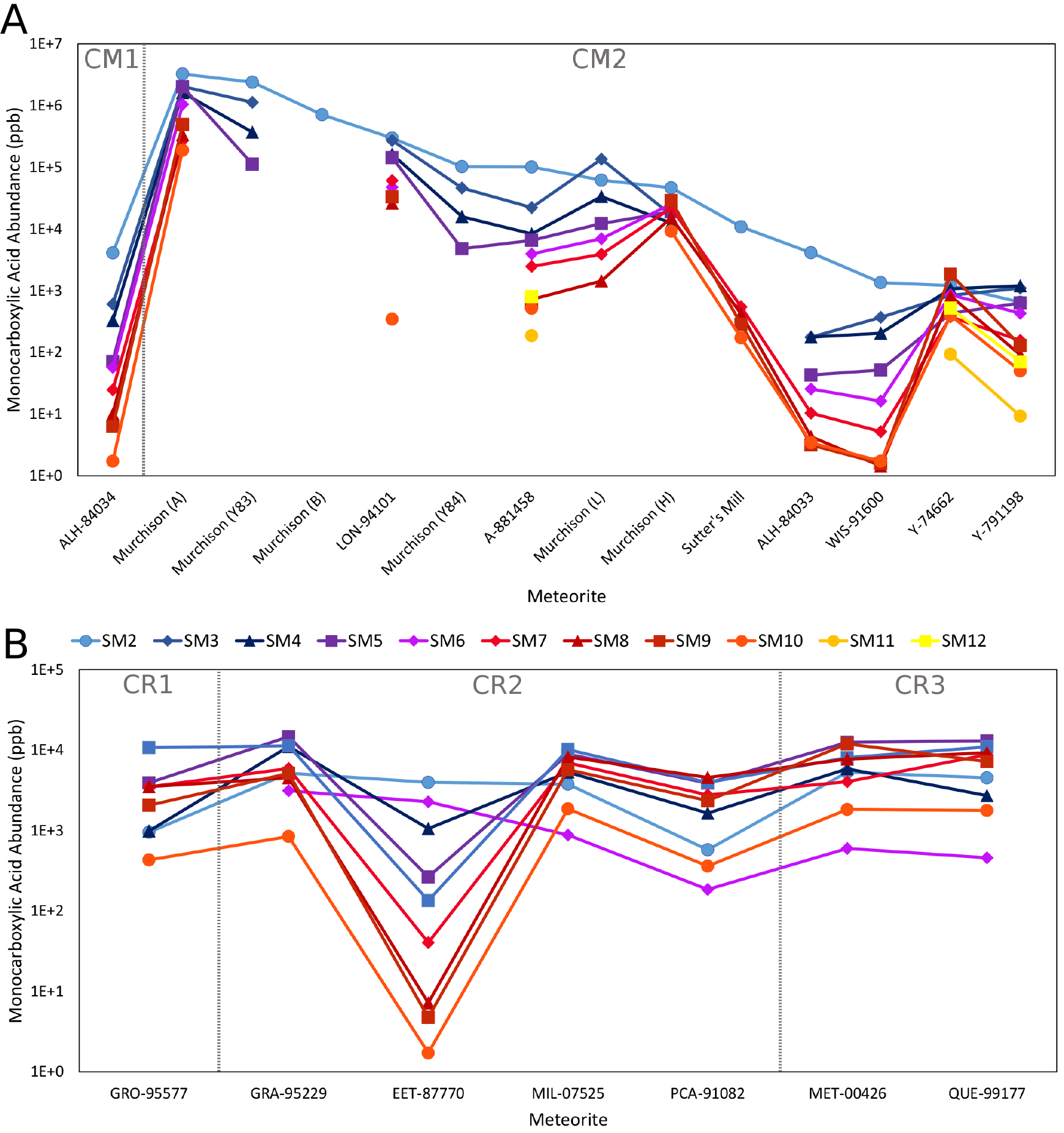}
	\caption{Abundances of straight-chain monocarboxylic acids (SCMA) in {\bf (A)} 14 CM-type and {\bf (B)} 7 CR-type meteorites, expressed in parts-per-billion (ppb). SCMA are labelled by their carbon chain length; see Table~\ref{tbl-abbr} for corresponding IUPAC names. Fatty acid abundances in CM and CR meteorites range from 1 ppb to 3.2$\times$10$^{6}$ ppb and 1 ppb to 1.5$\times$10$^{4}$ ppb, respectively.\label{Fig2}}
\end{figure*}

\begin{figure*}
	\includegraphics[width=0.8\linewidth]{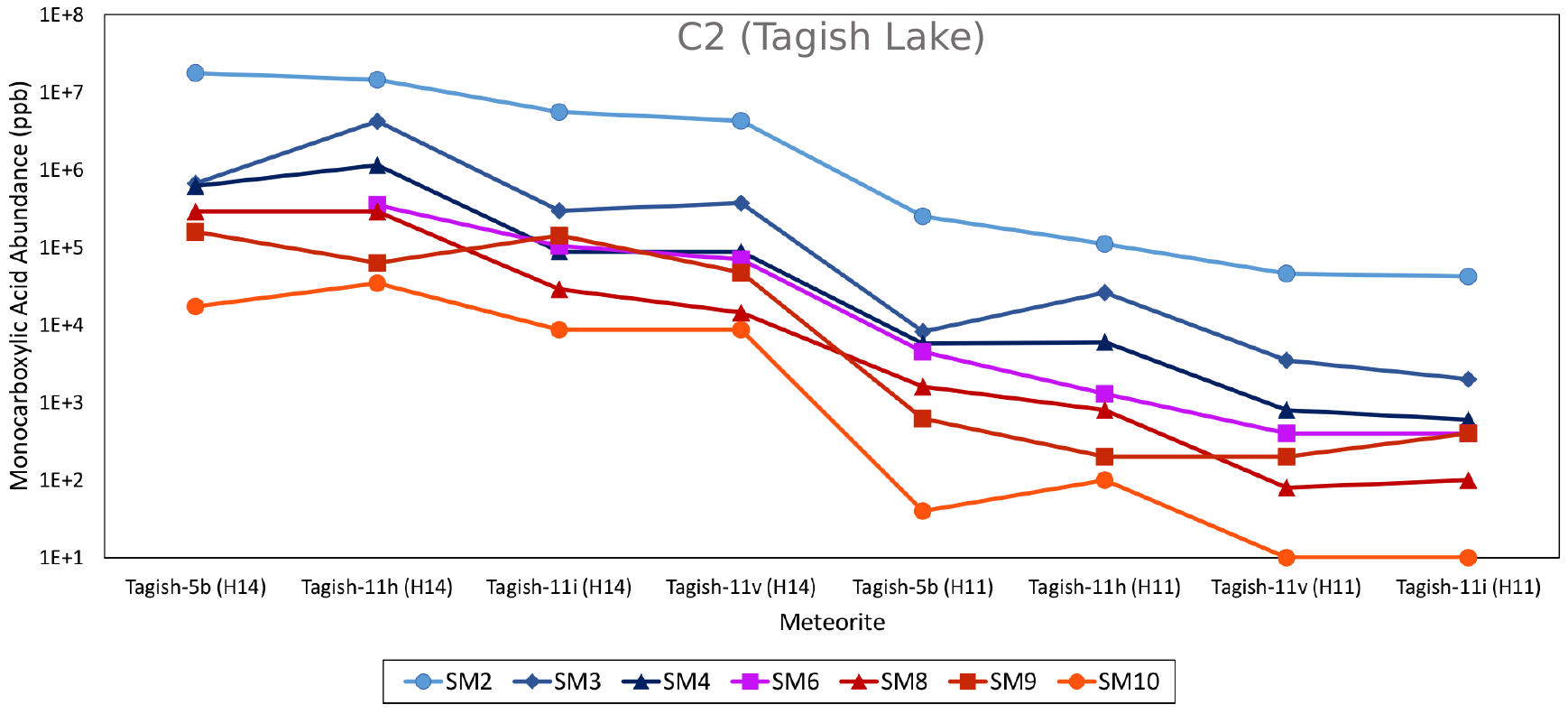}
	\caption{Abundances of straight-chain monocarboxylic acids (SCMA) in 8 samples of the Tagish Lake meteorite---an unclassified C-type carbonaceous chondrite---expressed in parts-per-billion (ppb). SCMA are labelled by their carbon chain length; see Table~\ref{tbl-abbr} for corresponding IUPAC names. Abundances range from 10 ppb to 1.8$\times$10$^{7}$ ppb. \label{Fig3}}
\end{figure*}

Fatty acid abundances are reported in parts-per-billion (ppb) per mass of bulk meteorite material. In Figures~\ref{Fig2}A, \ref{Fig2}B, and \ref{Fig3}, we display SCMA abundances in CM, CR, and Tagish Lake (ungrouped) meteorites, respectively. Data points are connected to aid in the tracing of abundances of any particular SCMA across different meteorites. SCMA are referred to by their carbon chain length (see the left column of Table~\ref{tbl-abbr}). Meteorites are ordered by petrological type, and then by decreasing abundance of SM2: the shortest and typically most abundant SCMA.

\begin{table}
	\caption{Shorthand names used for straight-chain monocarboxylic and dicarboxylic acids in this paper.\label{tbl-abbr}}
	\centering
	\begin{tabular}{clcl}
		Shorthand & IUPAC Name & Shorthand & IUPAC Name\\
		\hline \\[-2.5mm]
		SM2 &Ethanoic acid & SD2 & Ethanedioic acid\\
		SM3 &Propanoic acid & SD3 & Propanedioic acid\\
		SM4 &Butanoic acid & SD4 & Butanedioic acid\\
		SM5 &Pentanoic acid & SD5 & Pentanedioic acid\\
		SM6 &Hexanoic acid & SD6 & Hexanedioic acid\\
		SM7 &Heptanoic acid & SD7 & Heptanedioic acid\\
		SM8 &Octanoic acid & SD8 & Octanedioic acid\\
		SM9 &Nonanoic acid & SD9 & Nonanedioic acid\\
		SM10 &Decanoic acid & SD10 & Decanedioic acid\\
		SM11 &Undecanoic acid & SD11 & Undecanedioic acid\\
		SM12 &Dodecanoic acid & &\\ 
		\tableline
		
	\end{tabular}
	
	
\end{table}

One of the CM meteorites analyzed for SCMA is of petrologic type 1, and thirteen are of petrologic type 2 (see Appendix A for details on petrologic types). Half of the CM2 meteorites analyzed are samples of the Murchison meteorite; we have labeled these with identifiers in parentheses, as noted in Table~\ref{tbl-data}. SCMA with chain lengths 2--12 carbons long were detected. In most cases, SM2 is the most abundant SCMA within CM meteorites, with the other SCMA decreasing in abundance with increasing chain length. SM2--SM12 abundances in CM meteorites range from 1 ppb to 3.2$\times$10$^{6}$ ppb. Within just the Murchison meteorite, SM2 abundances span 2 orders of magnitude. This is consistent with other organics (e.g., amino acids, nucleobases) in samples of the Murchison meteorite, which also span at least an order of magnitude \citep{2014ApJ...783..140C,Reference46}.

No SCMA with chain lengths longer than 12 were detected. This is likely a consequence of the decreasing thermodynamic favourability of producing fatty acids in meteorite parent bodies with increasing chain lengths. In this case, SM13 abundances in meteorites may simply be too low to detect.

One of the CR meteorites analyzed for SCMA is of petrologic type 1, four are of petrologic type 2 and two are of petrologic type 3. The longest SCMA chain detected in CR meteorites is 10 carbons long. Only EET-87770 matches the trend of the CM meteorites, showing decreasing abundance with increasing chain length. EET-87770 also contains 1--2 orders of magnitude less SM5--SM10 compared to the other CR meteorites. All other CR meteorites contain higher abundances of SM4--SM9 compared to the shorter and longer SCMA. SCMA abundances in CR meteorites range from 1 ppb to 1.5$\times$10$^{4}$ppb.


SCMA abundances were analyzed in 8 samples of the ungrouped Tagish Lake meteorite, which is of petrologic type 2. The analyses performed did not determine SM5 and SM7 abundances, and did not search for fatty acids longer than SM10. SCMA abundances in the Tagish Lake meteorite samples follow the same pattern as the CM meteorites: SCMA abundances decline with increasing chain length. SCMA abundances in the ungrouped Tagish meteorites range from 10 ppb to 1.8$\times$10$^{7}$ ppb.

\begin{figure*}
	\includegraphics[width=0.8\linewidth]{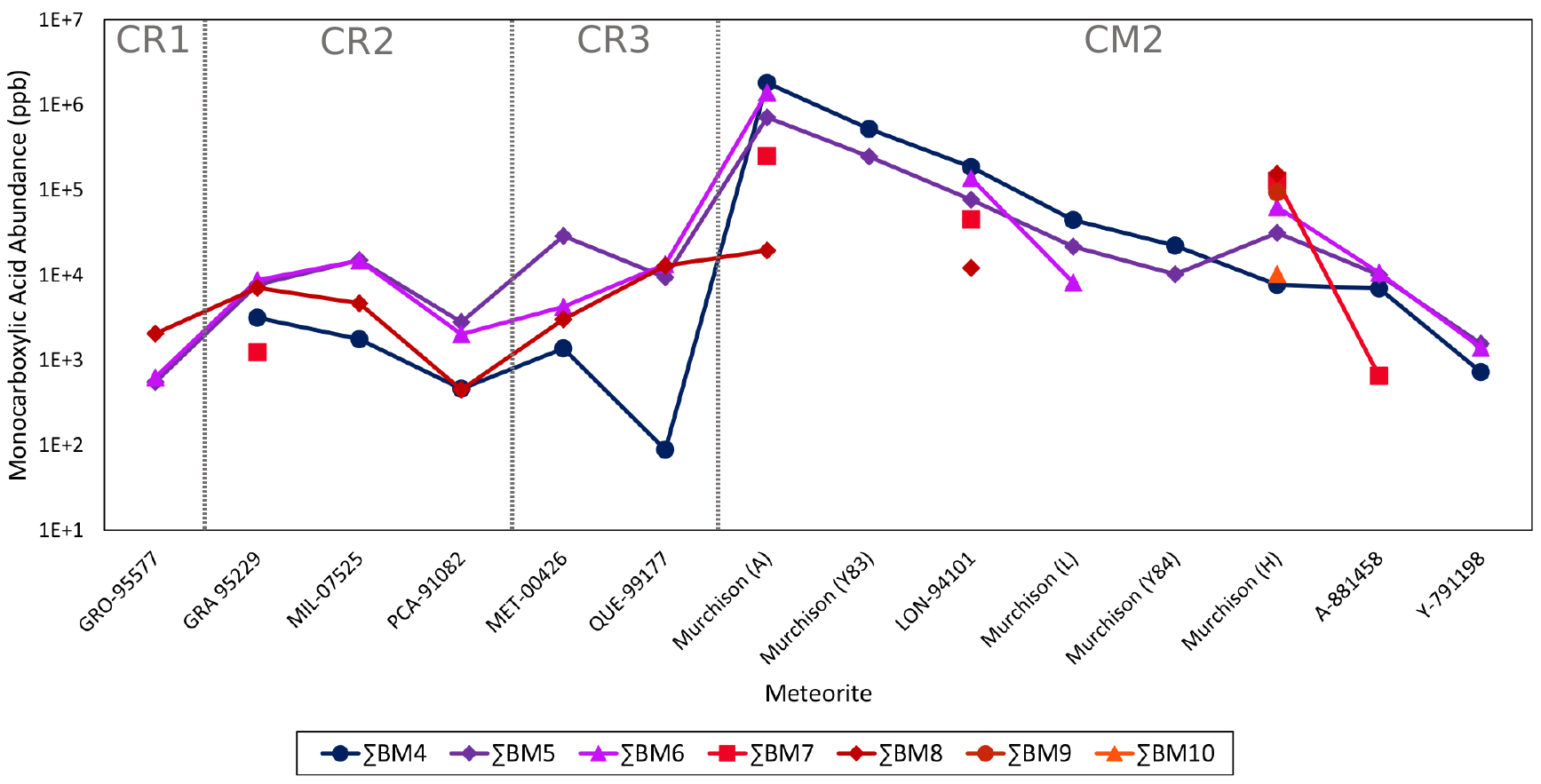}
	\caption{Abundances of branched-chain monocarboxylic acids (BCMA) in 6 CR-type and 8 CM-type meteorites, expressed in parts-per-billion (ppb). BCMA are labelled as a sum of BCMA with equivalent carbon numbers, e.g., $\Sigma$BM5 = S-(+)-2-methylbutyric acid, R-(-)-2-methylbutyric acid, 3-methylbutyric acid and 2,2-dimethylpropanoic acid. Abundances range from 1 ppb to 1.8$\times$10$^{6}$ ppb. \label{BCMA}}
\end{figure*}

For each number of carbons, there can be many different BCMA molecules. BCMA with the same number of carbons differ by the location in the carbon chain where branching occurs. For example, BCMA with 5 carbons include S-(+)-2-methylbutyric acid, R-(-)-2-methylbutyric acid, 3-methylbutyric acid and 2,2-dimethylpropanoic acid. We display BCMA as a sum of all permutations with the same number of carbons (e.g., $\Sigma$BM5 refers to the sum of the 4 molecules listed above) as some studies did not analyze these molecules separately. 

In Figure~\ref{BCMA}, we display BCMA abundances in CR- and CM-type meteorites. BCMA with 4--10 carbons were detected. CR meteorites have higher abundances of higher-carbon-number BCMA (i.e., $\Sigma$BM5, $\Sigma$BM6, $\Sigma$BM8) than the lowest-carbon-number BCMA (i.e., $\Sigma$BM4). BCMA in CM meteorites, however, demonstrate lower abundances with increasing carbon number. These behaviors match those of the SCMA in CR and CM meteorites. BCMA in CR and CM meteorites range from 80 ppb to 2.9$\times$10$^{4}$ ppb, and 520 ppb to 1.8$\times$10$^{6}$ ppb, respectively.

\begin{figure*}
	\includegraphics[width=0.8\linewidth]{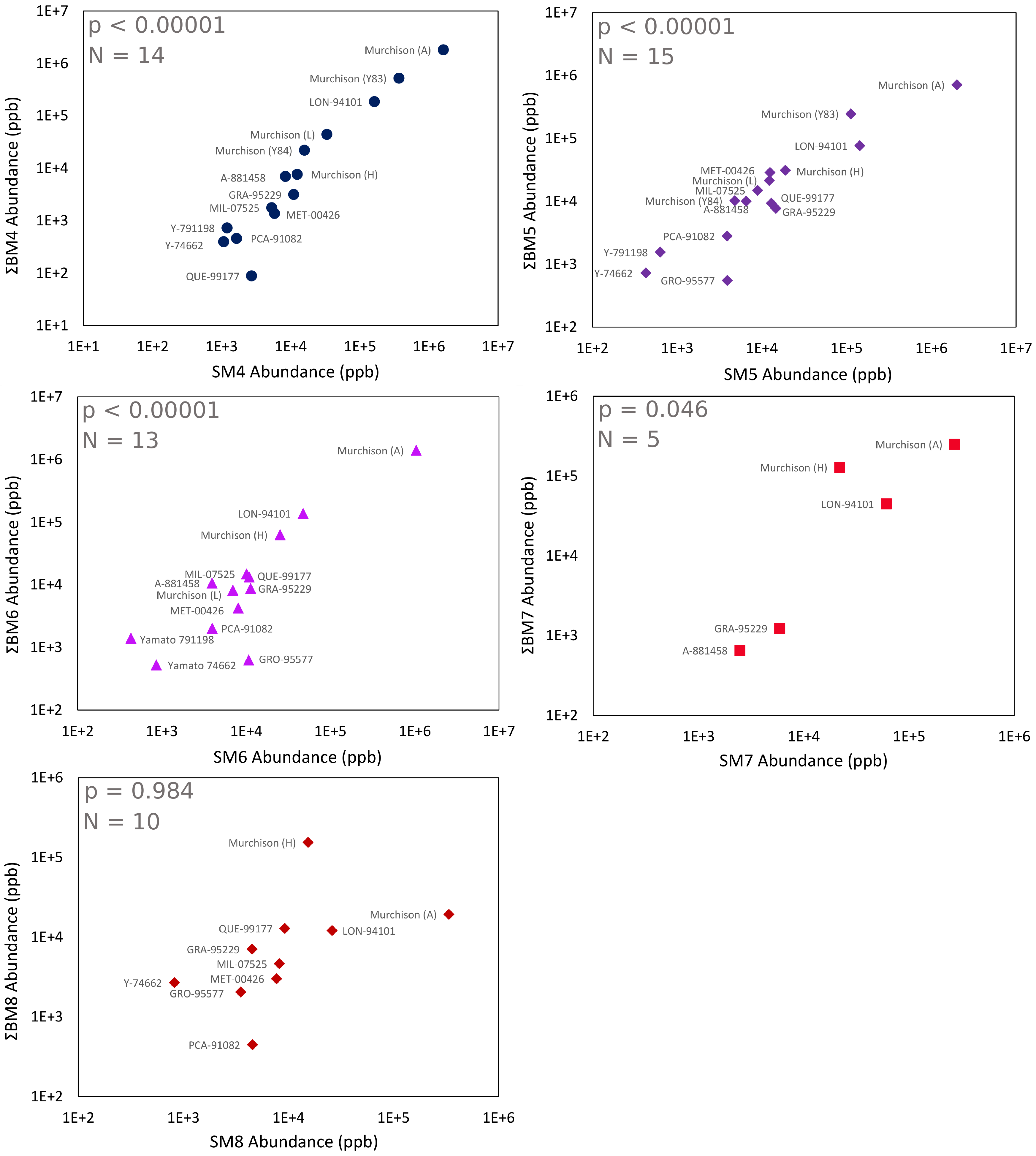}
	\caption{Correlation plots of meteoritic straight-chain monocarboxylic acids (SCMA) to the sum of meteoritic branched-chain monocarboxylic acids (BCMA) with the same carbon number. p-values correspond to the significance of correlation, and N is the sample size. For carbon numbers $\leq$ 7 (the top 4 frames), SCMA abundances are correlated to the sum of BCMA abundances with a significance of p $<$ 0.05. SCMA are labelled by their carbon chain length; see Table~\ref{tbl-abbr} for corresponding IUPAC names. BCMA are labelled as a sum of BCMA with equivalent carbon numbers, e.g., $\Sigma$BM5 = S-(+)-2-methylbutyric acid, R-(-)-2-methylbutyric acid, 3-methylbutyric acid and 2,2-dimethylpropanoic acid.  \label{compare1}}
\end{figure*}

\begin{figure*}
	\includegraphics[width=0.8\linewidth]{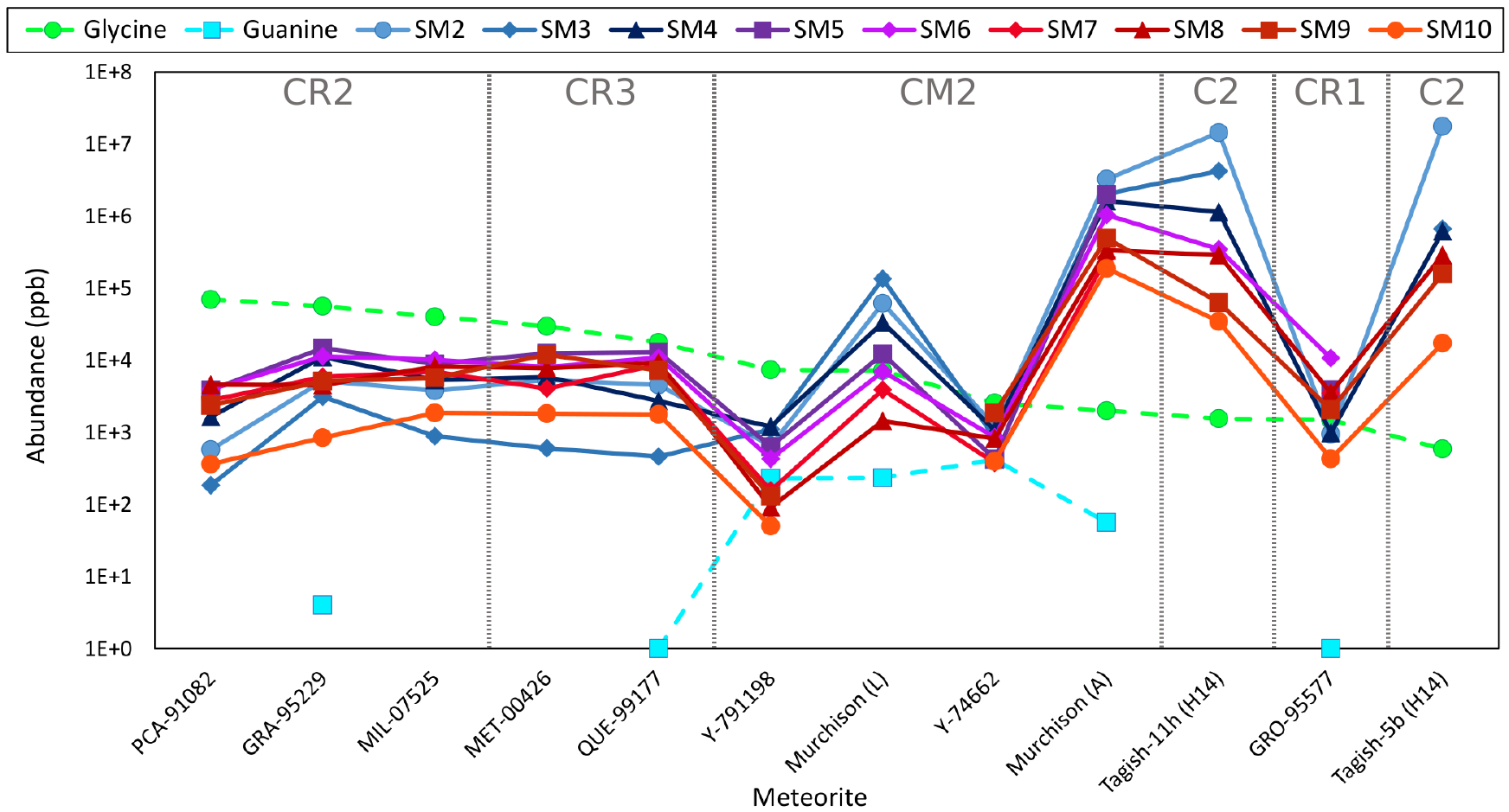}
	\caption{Straight-chain monocarboxylic acid (SCMA) abundances compared with glycine and guanine abundances in all meteorites with available data. For CR meteorites, all glycine and SCMA data were obtained from the same studies (i.e., the data came from the same meteorite samples). In all other cases, glycine and guanine data were selected from studies that obtained their meteorite samples from the same curator as the SCMA study (e.g., Smithsonian National Museum of Natural History, Arizona State University). The data sources are listed in Table~\ref{tbl-data}. Meteorites are sorted by monotonically decreasing glycine abundance. \label{compare2}}
\end{figure*}

For further meteoritic data, including dicarboxylic acids and the sum of all fatty acids within each meteorite, see Appendix D.

\subsection{Fatty Acid Comparisons and Correlations}

In Figure~\ref{compare1}, we compare the abundances of each SCMA with the sum of BCMA with the same carbon number (e.g. SM4 is compared with $\Sigma$BM4). We perform a Pearson correlation between each pair of abundances within all meteorites where measurements were available. For carbon numbers up to 7, the SCMA and sum of BCMA abundances are correlated with a significance of {\it p} $<$ 0.05 (however, for the SM7/$\Sigma$BM7 comparison, there were only 5 meteorite samples; therefore this particular correlation may not be statistically significant). 

For a carbon number of 8, the SCMA and sum of BCMA abundances are not correlated. This suggests that the dominant reaction mechanism forming SCMA up to $\sim$7 carbons in length in meteorite parent bodies may be the same as the reaction mechanism forming the BCMA with as many carbons. For a given carbon number $\lesssim$7, the abundances of SCMA and the sum of BCMA are very close across all meteorites. This suggests that it is equally favourable to form a particular SCMA with less than $\sim$7 carbons as it is to form all the BCMA with an equivalent number of carbons.

In Figure~\ref{compare2}, we compare SCMA abundances with the most abundant amino acid, glycine, and the most abundant nucleobase, guanine, in carbonaceous chondrites. In some studies, meteorite samples were analyzed for both SCMA and glycine (e.g., \citet{2014MPS...49..526H,2012PNAS..10911949P,2008PNAS..105.3700P}). In other cases, glycine, guanine and SCMA were analyzed in different samples of the same meteorite, in separate studies. For example, there are several large fragments of the Murchison meteorite which are curated at different facilities. For meteorites like Murchison, where glycine and guanine were analyzed in separate samples from the SCMA analysis, glycine and guanine abundances were collected from studies which obtained their meteorite samples from the same curator as the SCMA study (e.g. Smithsonian National Museum of Natural History). 

There does not appear to be a relation between the abundances of fatty acids and abundances of amino acids or nucleobases across meteorite samples. Glycine ranges from one order of magnitude more abundant to four orders of magnitude less abundant than SCMA in meteorites. Glycine is more abundant than all SCMA in the CR2, CR3 and two of the CM2 meteorites (Y-791198 and Y-74662). In the remaining CM2 meteorites, and the one CR1 meteorite, most or all SCMA are more abundant than glycine.

Why is there such a strong variation in which of these molecules are most abundant in carbonaceous chondrites? Part of the answer is likely due to the fact that SCMA and amino acids are produced from different reactants, i.e., the Strecker synthesis of amino acids requires NH$_3$ and HCN, whereas SCMA are likely produced from reactants like H$_2$, CO and other non-nitrogen-containing molecules (see Table~\ref{tbl-reactions}). The variable initial concentrations of these reactants within meteorite parent bodies will cause amino acid and SCMA abundances to fluctuate in meteorites. Competition between reactions also plays a role in how much of each molecule is produced. For example, if initial NH$_3$ abundances are low within a meteorite parent body, the formation of hydroxy acids will be favoured over amino acids \citep{Reference44}. 

Guanine abundances are comparable to 5 orders of magnitude lower than SCMA abundances in meteorites. Both nucleobases and SCMA can form via a Fischer-Tropsch-type (FTT) reaction, {\it i.e.}, 

\begin{equation}
\ce{CO + H2 + other ->C[metal catalyst]. products},
\end{equation}

However unlike SCMA, nucleobases also require NH$_3$ (in the place of ``other'' in the above equation) to form by FTT synthesis in meteorite parent bodies. Moreover, a significant portion of nucleobase synthesis in CR2 meteorite parent bodies is thought to take place through pathways involving HCN (e.g. \ce{5HCN + H2O -> Guanine + H2}) \citep{Reference106}. Thus, it is conceivable that the variation in relative guanine and SCMA abundances in meteorites can be explained by the difference in initial concentrations of these reactive nitrogen molecules within meteorite parent bodies.


In Figure~\ref{Aminocomp}, we compare amino acids and their monocarboxylic acid decay products within several meteorites and find no correlation.

\subsection{Fatty Acid Abundance Patterns}

\begin{figure*}
	\includegraphics[width=0.8\linewidth]{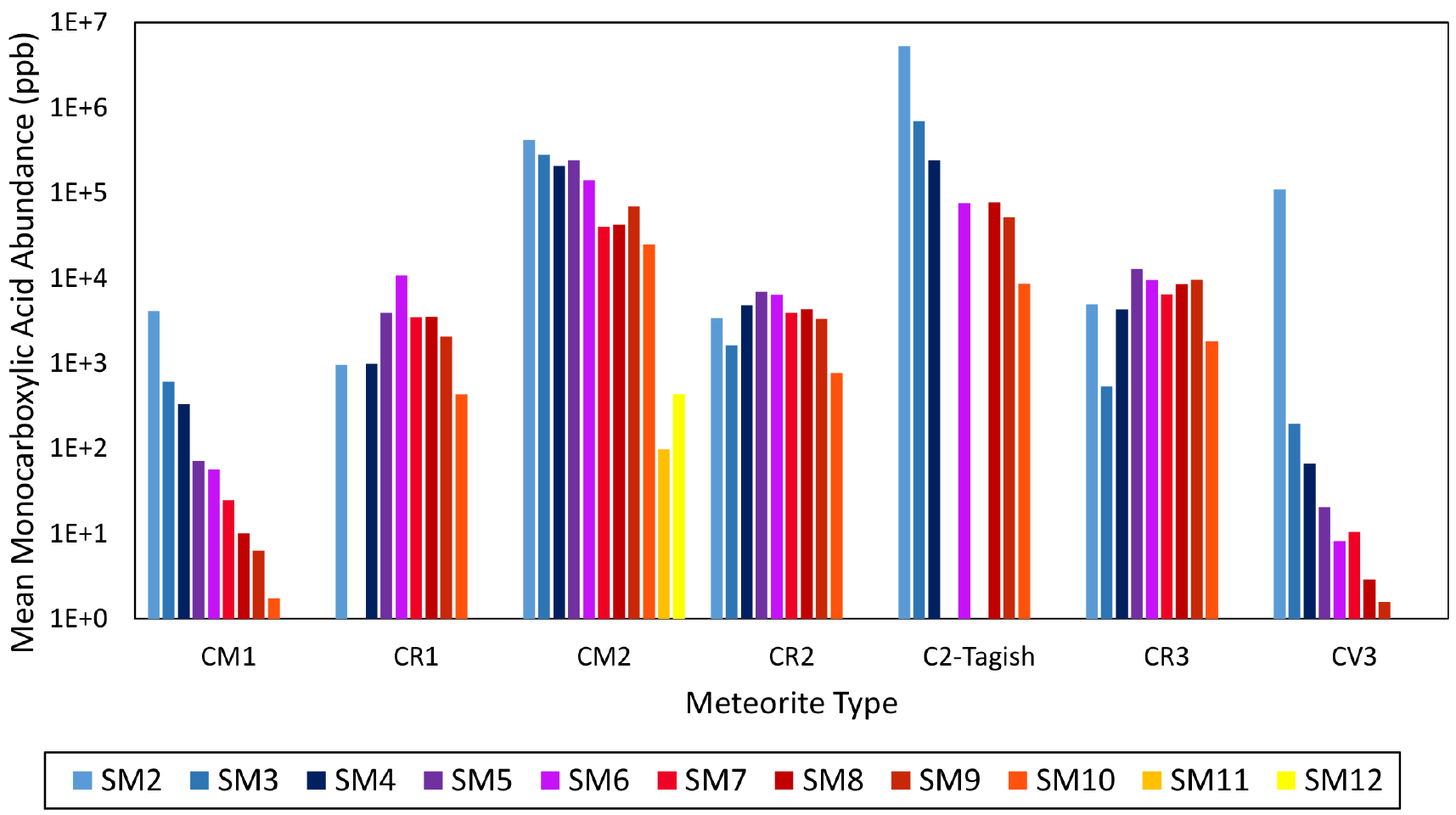}
	\caption{Average meteoritic straight-chain monocarboxylic acid (SCMA) abundances in each meteorite subclass. Average SCMA abundances are listed in order of increasing carbon chain length. \label{barplot}}
\end{figure*}

In Figure~\ref{barplot}, we plot the average SCMA abundances in the different meteorite subclasses. A downward trend of average SCMA abundances with increasing carbon length is apparent in the CM1, CM2, C2 (ungrouped), and CV3 meteorites. This trend is not apparent in the CR meteorites. The average SCMA abundances in CR meteorites peak at SM5 or SM6, and decrease with both increasing and decreasing carbon number from this central bump.

\begin{figure*}
	\includegraphics[width=0.8\linewidth]{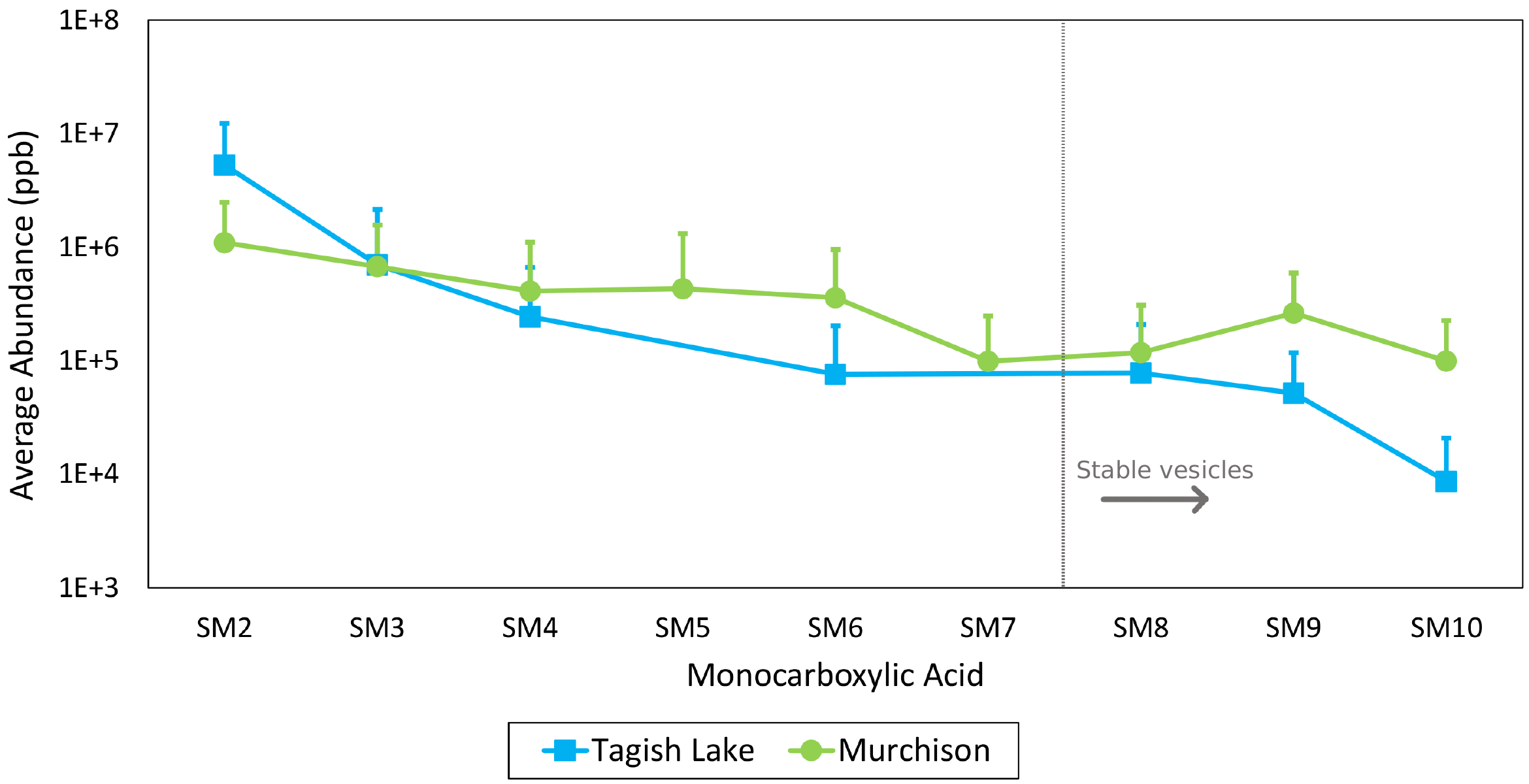}
	\caption{Average abundances of straight-chain monocarboxylic acids (SCMA) in the Tagish Lake and Murchison meteorites, expressed in parts-per-billion (ppb). Averages are calculated from all 8 samples of the Tagish Lake meteorite, and all 6 samples of the Murchison meteorite. Error bars represent 1 standard deviation, lower errorbars (not shown) extend to zero. SCMA are labelled by their carbon chain length; see Table~\ref{tbl-abbr} for corresponding IUPAC names. Both the Murchison and Tagish Lake meteorites exhibit decreasing average abundances of SCMA with increasing carbon chain length. SCMA with at least 8 carbons can form stable vesicles \citep{Reference429}. \label{trend}}
\end{figure*}

In Figure~\ref{trend}, we display the average abundance of each SCMA within the Murchison and Tagish Lake meteorites versus carbon-length. To calculate the averages for the Tagish Lake meteorite, all eight available meteorite sample analyses were used. For the SCMA averages in the Murchison meteorite, only the two samples which analyzed all SM2--SM10 monocarboxylic acids were used. 

We see a striking, simple pattern: in both the Tagish Lake and Murchison meteorites, average SCMA abundances decrease with increasing carbon number. This is likely a consequence of thermodynamics in the parent bodies, where larger chains are less energetically favourable to form in comparison to shorter chains, e.g., \citet{2009AsBio...9..483H}. Given that the Tagish Lake sample can be regarded as ``pristine'' in view of how it was collected, this figure also suggests that the Murchison data, in following the Tagish data trend, has relatively low levels of terrestrial contamination (see Appendix C). 

\begin{figure*}[hbtp]
	\includegraphics[width=0.8\linewidth]{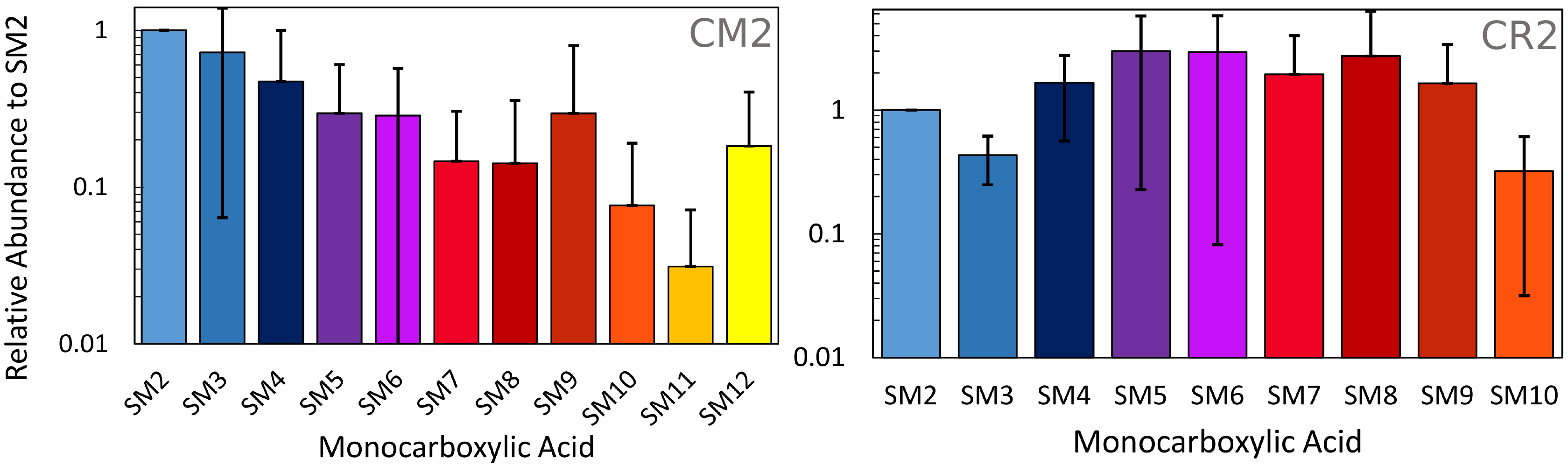}
	\caption{Mean abundances of straight-chain monocarboxylic acids (SCMA) in CM2 (left) and CR2 (right) meteorites relative to ethanoic acid (SM2). SCMA are labelled by their carbon chain length; see Table~\ref{tbl-abbr} for corresponding IUPAC names. Error bars represent 1 standard deviation. Lower error bars not shown if they extend to 0. \label{relativeAb}}
\end{figure*}

It is worth emphasizing that only SCMA with $\geq$ 8 carbons form stable vesicles \citep{Reference429}. Abundances of SCMA with $\geq$ 8 carbons are on average 0--2 orders of magnitude less abundant than those with $<$ 7 carbons in the Murchison and Tagish Lake meteorites. However, average abundances of SCMA with $\geq$ 8 carbons are still relatively high ($\sim$10$^4$--10$^5$ ppb) compared to biomolecules such as guanine ($\sim$10$^2$ ppb) and glycine ($\sim$10$^3$--10$^4$ ppb) in the same meteorites.

\subsection{Average Relative Fatty Acids Abundances}

Fatty acid abundances can differ by several orders of magnitude from one meteorite to another; therefore it is important to analyze the abundances of fatty acids in a relative way.
In the left and right panels in Figure~\ref{relativeAb}, we display the average abundances of SCMA relative to SM2 in CM2, and CR2 meteorites, respectively. In CM2 meteorites, average relative SCMA abundances generally exhibit a decline with increasing chain length, however SM12 is noticeably more abundant than SM10 and SM11. In CR2 meteorites, relative SCMA abundances peak around SM5 and SM6 and decline for increasing and decreasing carbons from that peak. Average relative SCMA abundances remain within 2 orders of magnitude of SM2 for CM2 meteorites, and within 1 order of magnitude of SM2 for CR2 meteorites.


\section{Abiotic Fatty Acid Synthesis}\label{sec3}

\begin{table*}[t]
\centering
	\caption{Candidate list for fatty acid synthesis within meteorite parent bodies. Reactions 5, 6, and 7 represent reaction processes, and no attempt is made to balance them. \label{tbl-reactions}}
	\begin{tabular}{lllll}
	\\
\multicolumn{1}{c}{No.} &
\multicolumn{1}{c}{Type} &
\multicolumn{1}{c}{Reaction} &
\multicolumn{1}{c}{\begin{tabular}{@{}c@{}}Carboxylic Acid(s) \\ Produced\end{tabular}} &
\multicolumn{1}{c}{Source(s)}\\ \hline
		1 & FTT & \ce{{\it n}CO + (2{\it n}-2)H2 ->C[Fe/none] (CH2)_{n}O2 + ({\it n}-2)H2O}, {\it n} $>$ 1 & SCMA, BCMA & \citet{2001OLEB...31..103R}; \\
		& & & & \citet{2006EPSL.243...74M}; \\
		& & & & \citet{Reference407} \\
        2 & NC & \ce{{\it n}CO2_{(aq)} + (3{\it n}-2)H2_{(aq)} -> (CH2)_{n}O2_{(aq)} + (2{\it n}-2)H2O} & SCMA & \citet{Reference414} \\
        3 & NC & \ce{2CO2_{(aq)} + H2_{(aq)} -> SD2_{(aq)}} & SCDA & \citet{Reference414} \\
        4 & NC & \ce{\alpha-Amino acid -> Carboxylic acid + NH3} & SCMA, BCMA & \citet{Reference420} \\
        5 & NC & \ce{PAH_{(s)} + CO_2_{(g)} + H_2O_{(l)} -> Carboxylic acid_{(aq)}} & SCMA & \citet{1990Natur.343..728S} \\
		6 & CA & \ce{CO_{(aq)} + C2H2_{(aq)} ->[NiS] Carboxylic acid_{(aq)}} & SCMA & \citet{Reference417} \\
		7 & CA & \ce{Formamide_{(aq)} ->[Carbonaceous meteorite dust] Carboxylic acid_{(aq)}} & SCMA, SCDA & \citet{Reference418} \\
		8 & CA & \ce{(CH2)_{n}O2 + CO + 2H2 <->[Fe|Ni|other] (CH2)_{n+1}O2 + H2O} & SCMA, BCMA & \citet{Reference419}; \\
		& & & & \citet{Reference407} \\
		9 & CA & \ce{Methanol + CO ->[Ni] SM2} & SCMA & \citet{Reference437} \\
		10 & CA & \ce{CH4 + CO2 ->[Pt + Al_2O_3| Pd + C] SM2} & SCMA & \citet{Reference435} \\
		11 & CA & \ce{Ethanol + H2O ->[CuCr] SM2 + 2H2} & SCMA & \citet{Reference436} \\[0.5mm]
		\hline
		\multicolumn{5}{l}{\footnotesize FTT: Fischer-Tropsch-type, NC: Non-catalytic, CA: Catalytic.} \\
		\multicolumn{5}{l}{\footnotesize PAH: polycyclic aromatic hydrocarbon.} \\
		\multicolumn{5}{l}{\footnotesize SCMA: Straight-chain monocarboxylic acid, SCDA: Straight-chain dicarboxylic acid, BCMA: Branched-chain monocarboxylic acid.}
	\end{tabular}
\end{table*}


In this section, we turn our attention to identifying the potential reaction pathways which formed fatty acids in meteorite parent bodies. We first search the literature for all known SCMA, BCMA, SCDA, and BCDA reaction pathways (see Table~\ref{glossary} for glossary of terms). We then narrow these pathways down to a list of candidate reactions that potentially produced fatty acids in meteorite parent bodies. Fatty acid reaction pathways only make it onto the candidate list if A) they can occur in the environment of meteorite parent body interiors, and B) their reactants are found in the spectra of comets. The latter criterion is required as comets are generally thought to be the most pristine bodies in our solar system, roughly containing the same basic starting materials that were available within planetesimals \citep{2004M&PS...39.1577S,2008tnoc.book..165R,2011IAUS..280..288A,Reference106}.

\subsection{Candidate Reaction Pathways in Planetesimals}


In Table~\ref{tbl-reactions}, we summarize 11 candidate reaction pathways that potentially produced fatty acids in meteorite parent bodies. We also list the fatty acid types (SCMA, BCMA, or SCDA) that the reactions are known to produce. None of the reaction pathways are known to produce BCDA, however this may be because BCDA are not commonly analyzed in experiments. A slash (\verb|/|) is used to signify ``or''.

\subsection{Reaction Mechanisms}

We categorize the main reaction mechanisms that form straight- and branched-chain monocarboxylic and dicarboxylic acids into 5 groups below. We list these mechanisms in rough order of decreasing likelihood of occurrence within meteorite parent bodies.

Fischer-Tropsch-type (FTT) reactions have long been suggested as a potential mechanism that formed the fatty acids in meteorites \citep{1981coge.conf....1A}. To form SCMA and BCMA via this mechanism, carbon monoxide must first be adsorbed onto the surface of a catalyst like iron or nickel, and then react with H$_2$ and water \citep{Reference419,Reference407}. The fatty acid chain can then be elongated or branched off through further reactions with carbon monoxide and H$_2$, or shortened by reacting with water \citep{Reference419,Reference407}. FTT reactions can occur both in the gas and aqueous phases.

Several non-catalytic (NC) reaction mechanisms are known to form fatty acids. The three most likely to occur in meteorite parent bodies are the reaction of CO$_2$ with H$_2$ \citep{Reference414}, the decay of $\alpha$-amino acids by loss of their amine functional group (-NH$_2$) \citep{Reference420}, and the reaction of CO$_2$ with polycyclic aromatic hydrocarbons (PAH) \citep{1990Natur.343..728S}. Other NC reactions are less likely to occur in meteorite parent bodies because at least one of their reactants are absent in comets. These include the degradation of sugars \citep{Reference440}, the hydrolysis of compounds such as esters \citep{Reference423,Reference406}, reactions of ketones with enols \citep{Reference424}, Arndt-Eistert synthesis \citep{Reference421} and Gallagher-Hollander degradation \citep{Reference422}. 


Some catalytic (CA) mechanisms (besides FTT synthesis) also produce fatty acids. The four most likely to produce fatty acids longer than SM2 in planetesimals are the reaction of CO with acetylene (C$_2$H$_2$) in the presence of nickel sulfide \citep{Reference417}, the reaction of aqueous formamide in the presence of carbonaceous chondrite material \citep{Reference418}, and the elongation and shortening of carboxylic acids that are adsorbed onto a catalyst like Fe or Ni through the reaction with CO and H$_2$, and H$_2$O, respectively \citep{Reference419,Reference407}. \citet{2004GeCoA..68.2185C} also produced a wide array of straight-chain and branched-chain monocarboxylic acids by reacting nonane thiol with carbon monoxide in aqueous solution in the presence of transition metal sulfides. However, nonane thiol is not found in comets, and is thus unlikely to be a reactant available within meteorite parent bodies. Carboxylic acids can also be produced by hydrolyzing nitriles or oxidizing aldehydes in the presence of acid catalysts such as sulfonic acid, hydrochloric acid, and quaternary ammonium bisulfate \citep{Reference438,Reference439}.

There are several possible reaction mechanisms likely to produce SM2 (ethanoic acid) in planetesimals, including the carbonylation of methanol in the presence of nickel \citep{Reference437}, the oxidation of methane in the presence of an platinum/alumina or palladium/carbon catalyst \citep{Reference435}, and production from ethanol via the ``aldehyde-water shift'' reaction in the presence of copper-chromium \citep{Reference436}.

Miller-Urey experiments have produced monocarboxylic and dicarboxylic acids using simple reactants such as methane and nitrogen \citep{Reference415,Reference433}. 


Cosmic ray simulation experiments have produced carboxylic acids through exposure of water and carbon dioxide to electrons \citep{2010ApJ...725.1002K}. Both cosmic rays and ultraviolet radiation can also initiate gas-phase ion chemistry, which has been demonstrated to produce carboxylic acids \citep{2003MNRAS.339L...7B}. Finally, photodecarboxylation (with visible light) of $\alpha$-hydroxycarboxylic acids has also been shown to produce carboxylic acids in the presence of an iodine catalyst \citep{Reference409}.	

The interior of a planetesimal is likely not an environment where cosmic rays, electron beams, or ultraviolet or visible light would be present; therefore we only consider FTT, NC and CA reactions as likely producers of carboxylic acids within meteorite parent bodies. We acknowledge that some meteoritic carboxylic acids may have originated from cosmic ray- and UV radiation-driven chemistry on the surfaces of icy grains that later accreted onto planetesimals; however we do not analyze this hypothesis in this paper.

\section{Suggested Dominant Reaction Pathways}\label{sec4}

\citet{Reference106} showed that the production of nucleobases in meteorite parent bodies is mainly driven by the initial concentration of each reaction's limiting reagent. In other words, whether a nucleobase reaction will dominate in a meteorite parent body is more dependent on the initial concentration of its reactants than the magnitude of its thermodynamic favourability. If this also holds true for fatty acid synthesis in meteorite parent bodies, then we can analyze which reaction pathways in Table~\ref{tbl-reactions} will dominate fatty acid synthesis in planetesimals based on the concentrations of their reactants within comets. Cometary concentrations are chosen instead of meteoritic concentrations as the former are likely the least aqueously altered bodies in the Solar System, and thus, their molecular inventory may better represent the interiors of early icy carbonaceous planetesimals \citep{2004M&PS...39.1577S,2008tnoc.book..165R,2011IAUS..280..288A,Reference106}. Furthermore, the mineralogy and chemistry of CI and CM meteorites is consistent with a cometary origin, and some carbonaceous asteroids are even known to exhibit cometary activity \citep{2003TrGeo...1..143S}.

In Table~\ref{tbl-comet}, we display cometary abundances of the reactants from the candidate reaction pathways of fatty acid synthesis within planetesimals. H$_2$O, CO, and H$_2$ are the most abundant fatty-acid-producing molecules in comets, likely being incorporated into these bodies via the agglomeration of condensates and/or clathrate hydrates in the protoplantary disk \citep{2010FaDi..147..509M}. (Clathrates are cage-like ice structures which host molecules like CO and H$_2$ in the gas phase.) These molecules suggest FTT synthesis may dominate the production of SCMA and BCMA in meteorite parent bodies. 

This conclusion is in agreement with the correlation between SCMA and the sum of BCMA in meteorites for carbon numbers $\lesssim$ 7 (see Figure~\ref{compare1}), which suggests the dominant reaction pathway produces both of these fatty acid types in meteorite parent bodies. Since CO$_2$ is only 2--4 times less abundant than carbon monoxide, reaction no. 2 may be the 2nd largest producer of SCMA. Due to this relatively high concentration of CO$_2$ in comets, reaction no. 3 may be the leading producer of dicarboxylic acids in planetesimals. Because the forward reaction no. 8 can elongate any SCMA and BCMA produced within planetesimals, this reaction is likely very important in the production of longer chain fatty acids in these bodies. However, because of the relatively low amount of formic acid (SM1) in comets, the source of shorter chain fatty acids for forward reaction no. 8 is likely not directly from the protoplanetary disk. Methanol and methane are about an order of magnitude less abundant in comets than CO, suggesting reaction nos. 9 and 10 may have played a small role in producing SM2 within planetesimals. All other reactants are more than an order of magnitude less concentrated in comets, and thus likely only produced a limited fraction of the fatty acids we observe in meteorites.

\begin{table}[t]
\centering
	\caption{Cometary concentrations of the potential reactants of fatty acid synthesis, in percent normalized to water. With the exception of H$_2$, ethanol, glycine, and benzene, molecular concentrations represent those measured in the spectra of comet Hale-Bopp \citep{2011ARA&A..49..471M}. H$_2$ has been detected in comets by rotationally resolved molecular hydrogen transitions, and its abundance is thought to be significant \citep{Reference21}. The ethanol concentration represents the value measured in the spectra of Comet Lovejoy \citep{2015SciA....115863B}. Glycine (the simplest amino acid) and benzene (the simplest PAH) concentrations are measured in the coma and on the surface of comet 67P/Churyumov-Gerasimenko \citep{2016SciA....2E0285A,2015Sci...349b0673W}.\label{tbl-comet}}
	\begin{tabular}{ll}
	\\
\multicolumn{1}{l}{Reactant} &
\multicolumn{1}{l}{Concentration (mol X/mol H$_2$O)}\\ \hline \\[-2.5mm]
		H$_2$O & 100 \\
		CO & 12--23 \\
		H$_2$ & significant \\
		CO$_2$ & 6 \\
		Methanol & 2.4\\
		CH$_4$ & 1.5 \\
		C$_2$H$_2$ & 0.1--0.3 \\
		Ethanol & 0.12\\
		Formic acid & 0.09 \\
		Formamide & 0.015 \\
		Glycine & 0--0.0025 \\
		Benzene & low \\[0.5mm]
		\hline
	\end{tabular}
\end{table}

\section{Differing Fatty Acid Abundance Patterns in CM and CR Meteorites}\label{sec5}

Why is there a discrepancy between the fatty acid abundance patterns in CR meteorites and those in CM or C (ungrouped) meteorites? One solution is that certain reaction mechanisms (e.g., non-catalytic (NC) reactions) could be active in different meteorite parent bodies. In previous work, we found that NC nucleobase reactions were likely more significant in CR parent bodies than in CM parent bodies, as carbon monoxide depletion would have reduced FTT synthesis in CR2 parent bodies \citep{Reference106}. The evidence for this is in the relatively high abundances of NH$_3$---the usual limiting reagent in the FTT synthesis of nucleobases---in CR2 meteorites \citep{2011GeCoA..75.7585M,2009GeCoA..73.2150P}. NH$_3$ would likely only have remained in these meteorites if chemical reactions depleted CO, the only carbon source in FTT synthesis, before NH$_3$ within CR2 parent bodies. 

Another explanation is contamination. Having a decreasing fatty acid abundance with an increasing number of carbon atoms is indicative of an extraterrestial origin (see Appendix C), which may suggest that the majority of the CR2 samples were subjected to some level of contamination.

The answer to the source of fatty acids in CR-type meteorites will remain unclear until more CR-meteorite samples are measured for fatty acids and analyzed for deuterium enrichment (see Appendix C for more details).

\section{Conclusions}\label{sec7}

In this paper, we compile and examine fatty acid abundances in meteorites and collect the potential reaction pathways responsible for their formation within the parent bodies. We focus our analysis on straight-chain monocarboxylic acids (SCMA), branched-chain monocarboxylic acids (BCMA), straight-chain dicarboxylic acids (SCDA) and branched-chain dicarboxylic acids (BCDA), as these molecules have been demonstrated to form vesicles or are suggested to be incorporated into the membranous boundary of the first protocells. Fatty acids have been detected in CM, CR, CV, and the Tagish Lake meteorites, spanning petrologic types 1--3. 

We discern various trends in the data by displaying fatty acid abundances in different ways, and comparing fatty acid abundances with those of amino acids, nucleobases and other fatty acids. We are especially interested in the systematic trends of abundances of fatty acids with increasing carbon number, and the relative abundances of straight versus branched chain molecules in view of their incorporation into vesicles upon mixing with water. We identified 11 candidate reaction pathways that potentially form fatty acids within meteorite parent bodies. We classified these pathways into three categories: Fischer-Tropsch-type (FTT), non-catalytic (NC), and catalytic (CA). We will follow up on these proposed reactions in a future theory paper on fatty acid synthesis.



The top three results from the meteoritic fatty acid analysis are:

\begin{itemize}
\item On average, CM2 and Tagish Lake (ungrouped) meteorites contain greater fatty acid abundances than other meteorite subclasses. Average abundances of SCMA in CM2 and Tagish Lake meteorites range from 10$^2$ ppb to 4$\times$10$^5$ ppb, and 10$^4$ ppb to 5$\times$10$^6$ ppb, respectively. Average abundances in CR2 meteorites range from 1--7$\times$10$^3$ ppb.
	\item Within CM, CV, and C (ungrouped) meteorites, SCMA abundances generally decrease with increasing chain length. Within CR meteorites, SCMA abundances peak at chain lengths 5--6 carbons long. These dissimilar patterns may suggest that the reaction pathways producing fatty acids in CR parent bodies differ from those that produce fatty acids in other carbonaceous chondrite parent bodies. Alternatively, the CR meteorite fatty acid abundance pattern may be the result of some level of contamination.
	\item For a given carbon number $\lesssim$7, SCMA abundances are correlated with the sum of BCMA abundances with a significance of {\it p} $<$ 0.05. This suggests that for shorter chains, the reaction pathway producing SCMA within meteorite parent bodies is related to that which produces BCMA, and for longer chains, different reaction pathways are involved.
\end{itemize}

Other significant results from analysis of the meteoritic data include:

\begin{itemize}
\item BCMA abundances similarly decrease with increasing chain length in CM meteorites, and peak in carbon numbers around 5--6 in CR meteorites.
	\item There is no correlation between SCMA abundances in meteorites and glycine or guanine abundances in meteorites. This is likely due to the fact that glycine and guanine are produced from NH$_3$ and/or HCN, which vary in initial concentrations in meteorite parent bodies, while SCMA are produced from molecules that do not contain nitrogen.
	\item There is no correlation between amino acids and their monocarboxylic acid decay products within meteorite parent bodies. This suggests the amino acid decay pathway (reaction no. 4 in Table~\ref{tbl-reactions}) is not the dominant source of monocarboxylic acids within meteorite parent bodies.
\end{itemize}

The top two results from the analysis of fatty acid reaction pathways are:

\begin{itemize}
	\item We identified 11 candidate abiotic reaction pathways which possibly produced fatty acids within carbonaceous chondrite parent bodies. However, only seven of these candidate reaction pathways were likely to produce fatty acids with more than 2 carbons. These reactions are categorized into three types: FTT, NC, and CA.
	\item FTT synthesis is likely the dominant reaction pathway producing SCMA and BCMA in meteorite parent bodies. This is due to the likely high initial concentrations of FTT reactants H$_2$O, CO, and H$_2$ in planetesimals.
\end{itemize}

Other significant results from analysis of fatty acid reaction pathways include:

\begin{itemize}
	\item NC reactions of H$_2$ with CO$_2$ likely produced an appreciable abundance of SCMA, BCMA, and may have been the most abundant producer of dicarboxylic acids within planetesimals.
	\item The elongation reaction was likely key in the production of longer chain SCMA and BCMA in meteorite parent bodies.
\end{itemize}

In a future study, we plan to computationally simulate each candidate reaction pathway in meteorite parent body environments, and compare the results to the measured meteoritic abundances collected in this paper. 







\acknowledgments

We would like to thank both anonymous referees for their comments which resulted in improvements to this paper. The research of J. C.-Y. L. was supported by an NSERC Undergraduate Student Research Award (USRA). B.K.D.P. is supported by an NSERC Postgraduate Scholarship-Doctoral (PGS-D). R.E.P. is supported by an NSERC Discovery Grant. D.L. was supported by an NSERC CREATE Canadian Astrobiology Training Program (CATP) Undergraduate Fellowship.

\bibliography{Carboxylic}





\beginappendixa

\section*{Appendix A - Meteorite Classification}

In this section, we briefly summarize the classification of carbonaceous chondrites.

Stony meteorites are split into one of two broad categories: chondrites and achondrites \citep{2006mess.book...19W}. The chondrites are undifferentiated (unmelted) and contain relatively pristine solar nebula material. These meteorites serve as a record of conditions during the formation of the Solar System. Chondrites have been further subdivided into carbonaceous, ordinary, and enstatite. It is the carbonaceous chondrites that are of interest in this study, as they are generally the only meteorites rich in organic content. It is thought that aqueous alteration and/or thermal metamorphism may have played a role in determining the organic content of carbonaceous chondrites \citep{1967GeCoA..31.1395H, 2002SGeo...23..411B}.

Carbonaceous chondrites are characterized by having $>$5 wt$\%$ of organic carbon \citep{2002SGeo...23..411B, 2005IAUS..231..479B, 2010M&PS...45.1948G}. They are also further subdivided into subclasses based on their mineralogical composition. These subclasses are based on a representative meteorite: CI (Ivuna-like), CM (Mighei-like), CR (Renazzo-like), CH (Allan Hills-like), CO (Ornans-like), CV (Vigarano-like), CK (Karoonda-like), and CB (Bencubbin-like). Meteorites of the same subclass are thought to originate from the same parent body. 

Meteorites are further assigned a petrological class from 1 to 6, indicating the extent to which they have been subjected to aqueous alteration or thermal metamorphism. Petrological class 1 indicates the greatest degree of aqueous alteration, while petrological class 6 indicates the highest degree of thermal metamorphism. Petrological class 3 indicates a meteorite that has undergone little to no alteration.

For a more detailed review of each meteorite subclass, we refer readers to one of our previous papers, \citet{2014ApJ...783..140C} or \citet{Reference44}. 

\beginappendixb

\section*{Appendix B - Analytical and Extraction Techniques for Fatty Acid Analyses in Meteorites}

\begin{table*}
	\caption{Summary of meteoritic fatty acid data sources. The glycine and guanine sources used in comparison analyses are also listed.\label{tbl-data}}
	\begin{tabular}{llllll}
		\footnotesize Type & \footnotesize Meteorite (Added Identifier) & \footnotesize \begin{tabular}{@{}c@{}}Analytical \\ Technique\end{tabular}& \footnotesize Fatty Acid Source & \footnotesize Glycine Source & \footnotesize Guanine Source\\
		\hline \\[-2.5mm]
		\footnotesize CM1 &\footnotesize Allan Hills (ALH) 84034 & \footnotesize GC-FID  &\footnotesize \citet{2011GeCoA..75.2309A} & &\\
		\footnotesize CR1 &\footnotesize Grosvenor Mountains (GRO) 95577 &\footnotesize GC-MS &\footnotesize \citet{2012PNAS..10911949P} &\footnotesize \citet{2012PNAS..10911949P} & \footnotesize \citet{2011PNAS..10813995C}\\
		\footnotesize CM2 &\footnotesize Allan Hills (ALGH) 84033  &\footnotesize GC-FID  &\footnotesize \citet{2011GeCoA..75.2309A} & & \\
		\footnotesize CM2 &\footnotesize Asuka (A) 881458 &\footnotesize GC-FID-MS &\footnotesize \citet{1999OLEB...29..187N} & & \\
		\footnotesize CM2 &\footnotesize Lone Wolf Nunataks (LON) 94101 &\footnotesize GC-FID &\footnotesize \citet{2014GeCoA.131....1A} & & \\
		\footnotesize CM2 &\footnotesize Murchison (Y84)  &\footnotesize GCC  &\footnotesize \citet{1984Natur.307..252Y} & & \\
		\footnotesize CM2 &\footnotesize Murchison (A) &\footnotesize GC-FID &\footnotesize \citet{2014GeCoA.131....1A} & \footnotesize \citet{2006MPS...41..889G} & \footnotesize \citet{2011PNAS..10813995C}\\
		\footnotesize CM2 &\footnotesize Murchison (B) &\footnotesize IEC &\footnotesize \citet{1993Metic..28..330B} & & \\
		\footnotesize CM2 &\footnotesize Murchison (H) &\footnotesize GC-FID &\footnotesize \citet{2005GeCoA..69.1073H} & & \\
		\footnotesize CM2 &\footnotesize Murchison (L) &\footnotesize GC-MS &\footnotesize \citet{1979Natur.282..396L} & \footnotesize \citet{1983AdSpR...3R...5C} & \footnotesize \citet{1979Natur.282..709S}\\
		\footnotesize CM2 &\footnotesize Murchison (Y83) &\footnotesize GCC &\footnotesize \citet{1983LPI....14..875Y} & & \\
		\footnotesize CM2 &\footnotesize Murchison (M) &\footnotesize GC-MS & \footnotesize \citet{2006MPS...41.1073M} & & \\
		\footnotesize CM2 &\footnotesize Sutter's Mill &\footnotesize GC-MS &\footnotesize \citet{2013LPI....44.2916D} & & \\
		\footnotesize CM2 &\footnotesize Wisconsin Range (WIS) 91600 &\footnotesize GC-FID &\footnotesize \citet{2011GeCoA..75.2309A} & & \\
		\footnotesize CM2 &\footnotesize Yamato (Y) 74662  &\footnotesize GC-FID-MS &\footnotesize \citet{Reference413} & \footnotesize \citet{1979Natur.282..394S} & \footnotesize \citet{Reference37}\\
		\footnotesize CM2 &\footnotesize Yamato (Y) 791198 &\footnotesize GC-FID-MS &\footnotesize \citet{Reference412} &\footnotesize \citet{Reference411} & \footnotesize \citet{Reference37}\\
		\footnotesize CR2 &\footnotesize Elephant Moraine (EET) 87770 &\footnotesize GC-MS &\footnotesize \citet{2011GeCoA..75.2309A} & & \\
		\footnotesize CR2 &\footnotesize Graves Nunataks (GRA) 95229 &\footnotesize GC-MS &\footnotesize \citet{2008PNAS..105.3700P} &\footnotesize \citet{2008PNAS..105.3700P} & \footnotesize \citet{2011PNAS..10813995C}\\
		\footnotesize CR2 &\footnotesize Miller Range (MIL) 07525 &\footnotesize GC-MS &\footnotesize \citet{2012PNAS..10911949P} &\footnotesize \citet{2012PNAS..10911949P} &\\
		\footnotesize CR2 &\footnotesize Pecora Escarpment (PCA) 91082 &\footnotesize GC-MS &\footnotesize \citet{2012PNAS..10911949P} &\footnotesize \citet{2012PNAS..10911949P} &\\
		\footnotesize C2 &\footnotesize Tagish Lake (P) &\footnotesize GC-MS &\footnotesize \citet{2002MPS...37..687P} & & \\
		\footnotesize C2 &\footnotesize Tagish Lake 5b (H11) &\footnotesize GC-MS &\footnotesize \citet{2011Sci...332.1304H} & & \\
		\footnotesize C2 &\footnotesize Tagish Lake 5b (H14) &\footnotesize GC-MS &\footnotesize \citet{2014MPS...49..526H} &\footnotesize \citet{2014MPS...49..526H} & \\
		\footnotesize C2 &\footnotesize Tagish Lake 11h (H11) &\footnotesize GC-MS &\footnotesize \citet{2011Sci...332.1304H} & & \\
		\footnotesize C2 &\footnotesize Tagish Lake 11h (H14) &G\footnotesize C-MS &\footnotesize \citet{2014MPS...49..526H} &\footnotesize \citet{2014MPS...49..526H} \\
		\footnotesize C2 &\footnotesize Tagish Lake 11i (H11) &\footnotesize GC-MS &\footnotesize \citet{2011Sci...332.1304H} & & \\
		\footnotesize C2 &\footnotesize Tagish Lake 11i (H14) &\footnotesize GC-MS &\footnotesize \citet{2014MPS...49..526H} & & \\
		\footnotesize C2 &\footnotesize Tagish Lake 11v (H11) &\footnotesize GC-MS &\footnotesize \citet{2011Sci...332.1304H} & & \\
		\footnotesize C2 &\footnotesize Tagish Lake 11v (H14) &\footnotesize GC-MS &\footnotesize \citet{2014MPS...49..526H} & & \\
		\footnotesize CR3 &\footnotesize Meteorite Hills (MET) 00426 &\footnotesize GC-MS &\footnotesize \citet{2012PNAS..10911949P} &\footnotesize \citet{2012PNAS..10911949P} & \\
		\footnotesize CR3 &\footnotesize Queen Alexandra Range (QUE) 99177 &\footnotesize GC-MS &\footnotesize \citet{2012PNAS..10911949P} &\footnotesize \citet{2012PNAS..10911949P} & \footnotesize \citet{2011PNAS..10813995C}\\
		\footnotesize CV3 &\footnotesize Allende &\footnotesize IEC &\footnotesize \citet{1993Metic..28..330B} & & \\
		\footnotesize CV3 &\footnotesize Meteorite Hills (MET) 00430 &\footnotesize GC-FID &\footnotesize \citet{2011GeCoA..75.2309A} & & \\		
		\tableline
		
	\end{tabular}
	
	
	
\end{table*}

In Table~\ref{tbl-data}, we present a summary of the fatty acid meteoritic data.
Most studies make use of gas chromatography (GC) to quantify fatty acid abundances, however this is usually coupled to a mass spectrometer (GC-MS), a flame ionization detector (GC-FID), both (GC-FID-MS) or a combustion system (GCC). GC separates the molecules in a mixture by dissolving the mixture in a gas and sending it through a column. The molecules travel at different speeds through the column, separating them for detection. If a FID is coupled to the system, the mixture is then sent through a flame (typically H$_2$), which oxidizes organics, producing ions for detection. If a MS is coupled to the system, the mixture is then bombarded with electrons and accelerated through an electric or a magnetic field. This separates ions based on their mass-to-charge ratio for detection. Finally, in the case of a GCC, the mixture is passed through an oven (typically a tube containing high-temperature wires), which oxidizes organics, producing ions for detection. The only additional technique used to analyze fatty acids in meteorites is ion-exclusion chromatography (IEC), which separates molecules in aqueous solution by passing them through a column containing an ion exchange resin, taking advantage of each species' varying tendency to ionize in solution.

Studies also varied in their extraction techniques (i.e. the initial separation of fatty acids from other compounds in a mixture prior to feeding them into the chromatograph). Some studies used liquid-liquid extractions, where molecules are separated based on their relative solubilities in two different solvents. Recent studies (e.g. \citet{2014MPS...49..526H,2014GeCoA.131....1A}) made use of solid-phase microextraction (SPME), in which fatty acids are gathered by a coated fibre before feeding them into the GC. By forgoing the need of a solvent, SPME reduces the evaporative loss of fatty acids compared to the former technique \citep{2005GeCoA..69.1073H}.

The study by \citet{1979Natur.282..396L} was the only one to include uncertainties with their fatty acid abundances, ranging from 1.6$\%$ to 33$\%$---with lower uncertainties corresponding to the shorter-chained fatty acids.


\beginappendixc

\section*{Appendix C - Possibility of Contamination}

In this section, we discuss the evidence for and against an extraterrestrial origin for the meteoritic fatty acids analyzed in this paper. At the end of this section, we use this evidence to assess the likelihood of contamination.

\subsection*{Supporting an Extraterrestial Origin}

The most scientifically rigorous method used to determine the origin of meteoritic fatty acids is to determine the deuterium enrichment in the sample \citep{2002MPS...37..687P,2005GeCoA..69.1073H,2006MPS...41.1073M,2011Sci...332.1304H,2011GeCoA..75.2309A,2014MPS...49..526H,2014GeCoA.131....1A}. Interstellar compounds exhibit an enrichment in deuterium as a result of low temperature ion-molecule reactions \citep{1987Natur.326..477E}. When it comes to biochemical reactions, on the other hand, the lighter isotope is very often enriched due to kinetic effects \citep{Reference441}. Hydrogen and deuterium have the largest mass difference relative to other elements and their isotopes, therefore they exhibit the largest range of isotope ratios, e.g. $\delta D \sim$ -150 to -25 $\permil$ for sedimentary rocks, $\delta D$ = 0 $\permil$ for ocean water, $\delta D \sim$ -350 to +50 $\permil$ for snow and rain water, and $\delta D \sim$ +100 to +3600 $\permil$ for interstellar and meteorite parent body sources \citep{Reference441,2008PNAS..105.3700P}. Carboxylic acids of extraterrestrial origin should therefore exhibit deuterium enrichment relative to terrestrial values \citep{1987Natur.326..477E,2003TrGeo...1..269G}.

Out of the 17 meteoritic analyses reviewed in this paper, six performed deuterium analyses and concluded that their samples were low in contamination \citep{2002MPS...37..687P,2005GeCoA..69.1073H,2011Sci...332.1304H,2011GeCoA..75.2309A,2014MPS...49..526H,2014GeCoA.131....1A}. These analyses include the two most widely studied meteorites for fatty acids: the Murchison and Tagish Lake meteorites. 

The Tagish Lake meteorite is well known to be the most pristine carbonaceous meteoritic sample collected to date. This is due to the fact that it was collected without direct hand contact just a few days after it fell, and all but one fragment have been kept below 0$^{\circ}$C since their collection \cite{2011Sci...332.1304H,2014MPS...49..526H}. 

\begin{figure*}[hbtp]
	\includegraphics[width=0.8\linewidth]{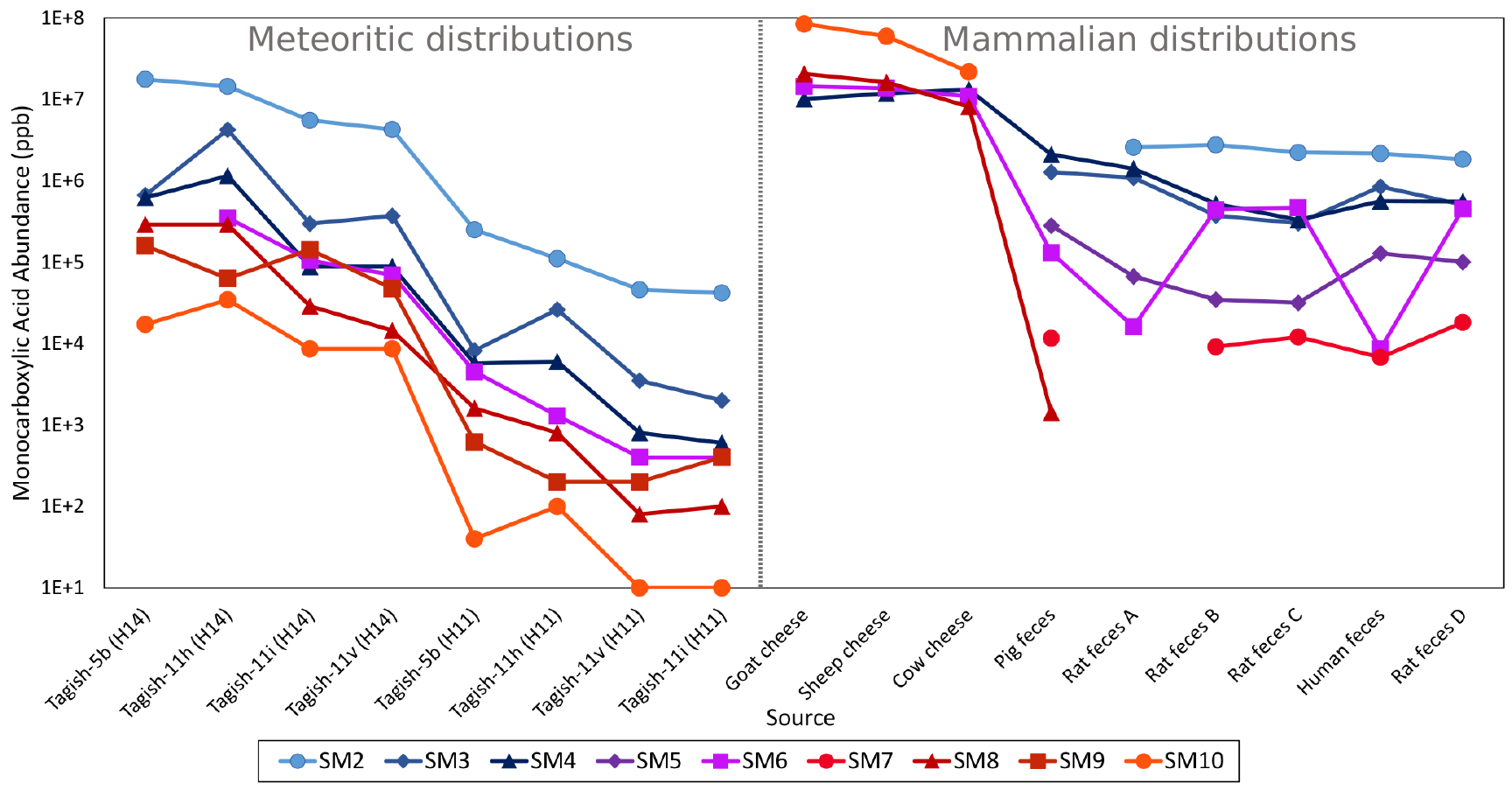}
	\caption{Comparion of straight-chain monocarboxylic acids (SCMA) abundances in Tagish Lake meteorite samples (left) to those in various mammalian sources (right). Data collected from a variety of sources: Tagish Lake samples \citep{2011Sci...332.1304H,2014MPS...49..526H}, cheese samples \cite{Reference443}, pig feces sample \citep{Reference445}, rat feces sample A \citep{Reference444}, and rat feces samples B, C, and D and human feces sample \citep{Reference446}. Studies varied in the fatty acids they analyzed, no complete analysis of SM2--SM10 in a biological sample was found. SCMA are labelled by their carbon chain length; see Table~\ref{tbl-abbr} for corresponding IUPAC names. \label{AppendixFig}}
\end{figure*}

It has been suggested that the decrease in monocarboxylic acid concentrations with increasing number of carbon atoms is indicative of extraterrestial origin, as living organisms are more selective in the fatty acids they produce and consume \citep{1979Natur.282..396L,Reference442}. In Figure~\ref{AppendixFig}, we compare the Tagish Lake fatty acid abundances with those from the small available subset of biological sources in the literature \citep{Reference443,Reference444,Reference445,Reference446}. 

In the Tagish Lake samples, SCMA abundances decline with increasing chain length, with only a few samples exhibiting slight variations to this trend. This pattern can not be equally stated for the biological (mammalian) samples. For example, in two of the three cheese samples, the exact opposite pattern is seen, and in all cheese samples SM10 is more abundant than the 3 shorter analyzed fatty acids. Furthermore, in the pig and four rat feces samples, SM4 is more abundant than SM3, and in three of the rat feces samples, SM6 is more abundant than SM5.

Hexadecanoic acid (SM16), and octadecanoic acid (SM18) are the dominant terrestrial monocarboxylic acids \citep{1969Natur.222..364H}. Neither of these fatty acids were reported in any of the meteoritic fatty acid analyses in this paper.

Finally, some of the meteoritic analyses suggested that the lack of contamination of amino acids in their meteorite samples suggests that the fatty acids in these samples are also of extraterrestrial origin \citep{Reference412,Reference413,2008PNAS..105.3700P,2012PNAS..10911949P,2013LPI....44.2916D}.

\subsection*{Opposing an Extraterrestial Origin}


\citet{2011GeCoA..75.2309A} note that shorter acids such as ethanoic acid (SM2) are susceptible to contamination during sample storage due to microbial activities. This may be the reason why SM2 exhibits the lowest enrichments in deuterium compared to the longer monocarboxylic acid chains \citep{2014MPS...49..526H}. Furthermore, three meteorite samples (Murchison (H), ALH-84034, and MET-00430) had two or three fatty acids exhibiting low negative $\delta D$ values \citep{2011GeCoA..75.2309A}, meaning these data may have been subjected to a small degree of contamination. 

The deuterium analyses for the Tagish Lake meteorite samples are incomplete. Deuterium isotopes were only measured for sample 5b, and only for fatty acids with 5 carbons or less. The deuterium enrichments of fatty acids in sample 5b suggest it was largely free of contamination, however sample 11v contained many contaminants from its storage medium, including terrestrial limonene \citep{2014MPS...49..526H}. 


\subsection*{Summary}

Meteoritic contamination is never completely zero, as all of the meteorite samples to date have been exposed to the Earth's atmosphere and surface for some duration. Contamination can, however, be low enough (e.g. within a few percent) to not greatly affect the organic abundances and distributions. Unfortunately, one cannot use isotopes to determine the percent contamination within meteorites without knowing the exact isotopic composition of the reservoir out of which the meteorite parent body was built. In any case, the general deuterium enrichment in several meteoritic samples suggests that contamination should be low, and should not greatly affect the fatty acid abundances and distributions.

The consistent decreasing fatty acid abundance for increasing carbon chain length for the CM and Tagish Lake meteorites is another strong indication that the fatty acids within these meteorites are extraterrestrial in origin.

As for the CR meteorites analyzed in this paper, with the exception of EET-87770, their fatty acid abundance patterns which peak around SM5 and SM6 and decline for increasing and decreasing carbons may hint that they have been subjected to some level of contamination. On the other hand, based on the pattern of fatty acids in the small subset of biological samples available in the literature (see Figure~\ref{AppendixFig}), we might expect contaminated samples to exhibit a less distinct pattern than the consistent pattern exhibited across all but one CR meteorite sample. Indeed, this distinct pattern seen in the CR meteorite samples may be the result of different reactions mechanisms being active in different meteorite parent bodies.

As a whole, there is a strong case for the extraterrestrial origin of all the CM and Tagish Lake meteoritic fatty acids. However, we advise caution in concluding the same origin for the CR meteoritic fatty acids until the deuterium isotope standard for contamination analysis is undertaken.

\beginappendixd

\section*{Appendix D - Additional Data and Comparisons}\label{additional}

\begin{figure*}
	\includegraphics[width=0.8\linewidth]{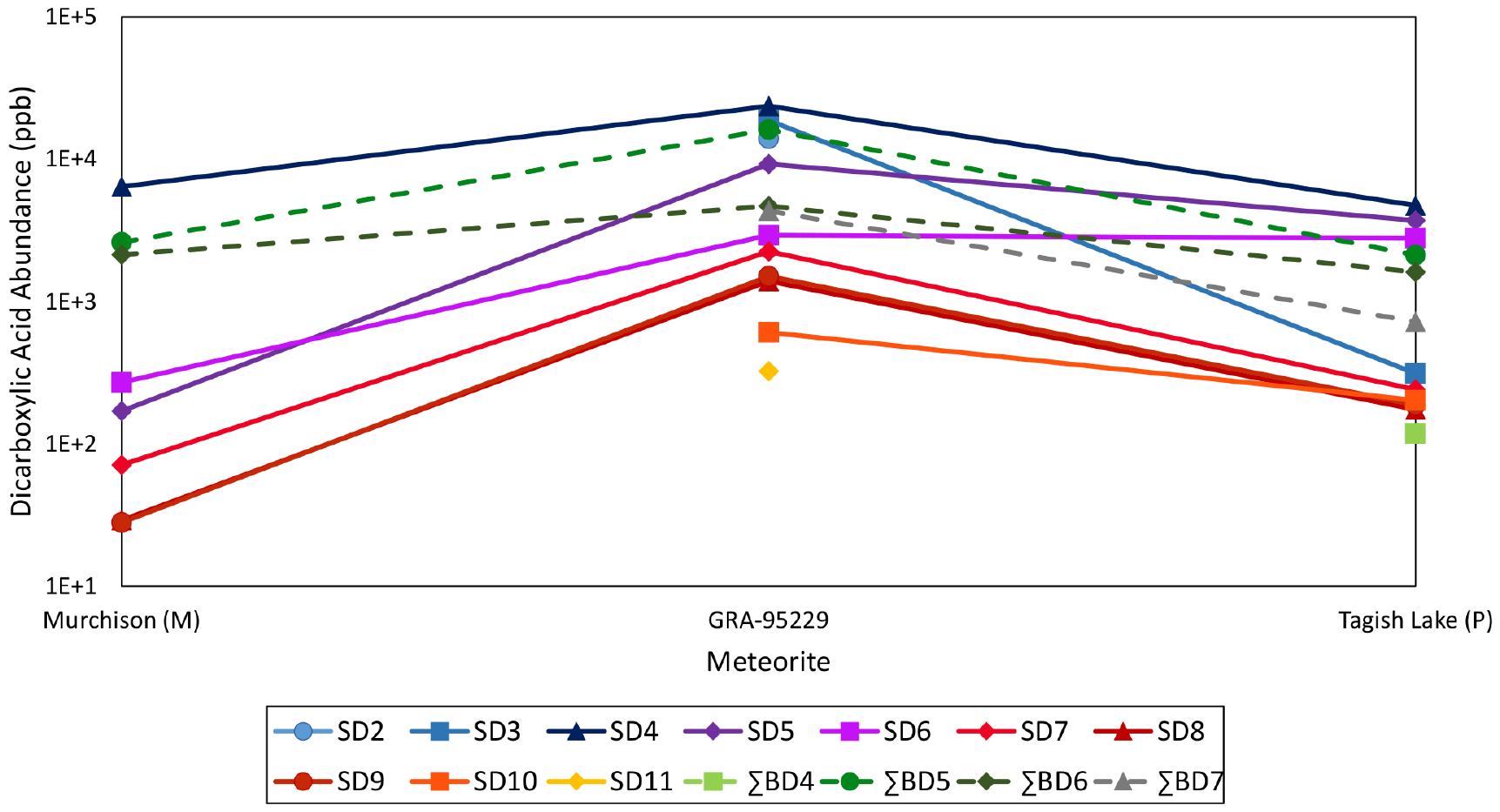}
	\caption{Abundances of straight-chain dicarboxylic acids (SCDA) and branched-chain dicarboxylic acids (BCDA) in 1 CM2 meteorite (Murchison), 1 CR2 meteorite (GRA-95229), and 1 C2 meteorite (Tagish Lake), expressed in parts-per-billion (ppb). SCDA are labelled by their carbon chain length; see Table~\ref{tbl-abbr} for corresponding IUPAC names. BCDA are labelled as the sum of BCDA with equivalent carbon numbers, e.g., $\Sigma$BD5 = methylbutanedioic acid and ethylpropanedioic acid. SCDA abundances and the sum of BCDA abundances range from 30 ppb to 2.4$\times$10$^{4}$ ppb, and 120 ppb to 1.6$\times$10$^{4}$ ppb, respectively. \label{dicarb}}
\end{figure*}

\subsection*{Meteoritic Dicarboxylic Acids}

In Figure~\ref{dicarb}, we display SCDA and the sum of BCDA abundances in the three meteorites analyzed for dicarboxylic acids to date. Similarly to monocarboxylic acids, dicarboxylic acids have several branched-chain forms for any given carbon number which differ based on the location in the carbon chain where branching occurs. Meteoritic data on dicarboxylic acids is limited to one CM2, one CR2, and one Tagish Lake (ungrouped) meteorite sample. SCDA with 2--11 carbons and BCDA with 4--7 carbons have been detected in these meteorites. SD4 is the most abundant dicarboxylic acid in all meteorites; however SD2 and SD3 are also very abundant in the one meteorite sample in which they were detected. 

Abundances of both SCDA and BCDA also generally decrease with increasing carbon number for all meteorite types; however we must be careful in drawing conclusions from a data set of this size. SCDA abundances in meteorites range from 30 ppb to 2.4$\times$10$^{4}$ ppb, and the sum of BCDA abundances range from 120 ppb to 1.6$\times$10$^{4}$ ppb.

\subsection*{Total Fatty Acids in Meteorites}

\begin{figure*}
	\includegraphics[width=0.8\linewidth]{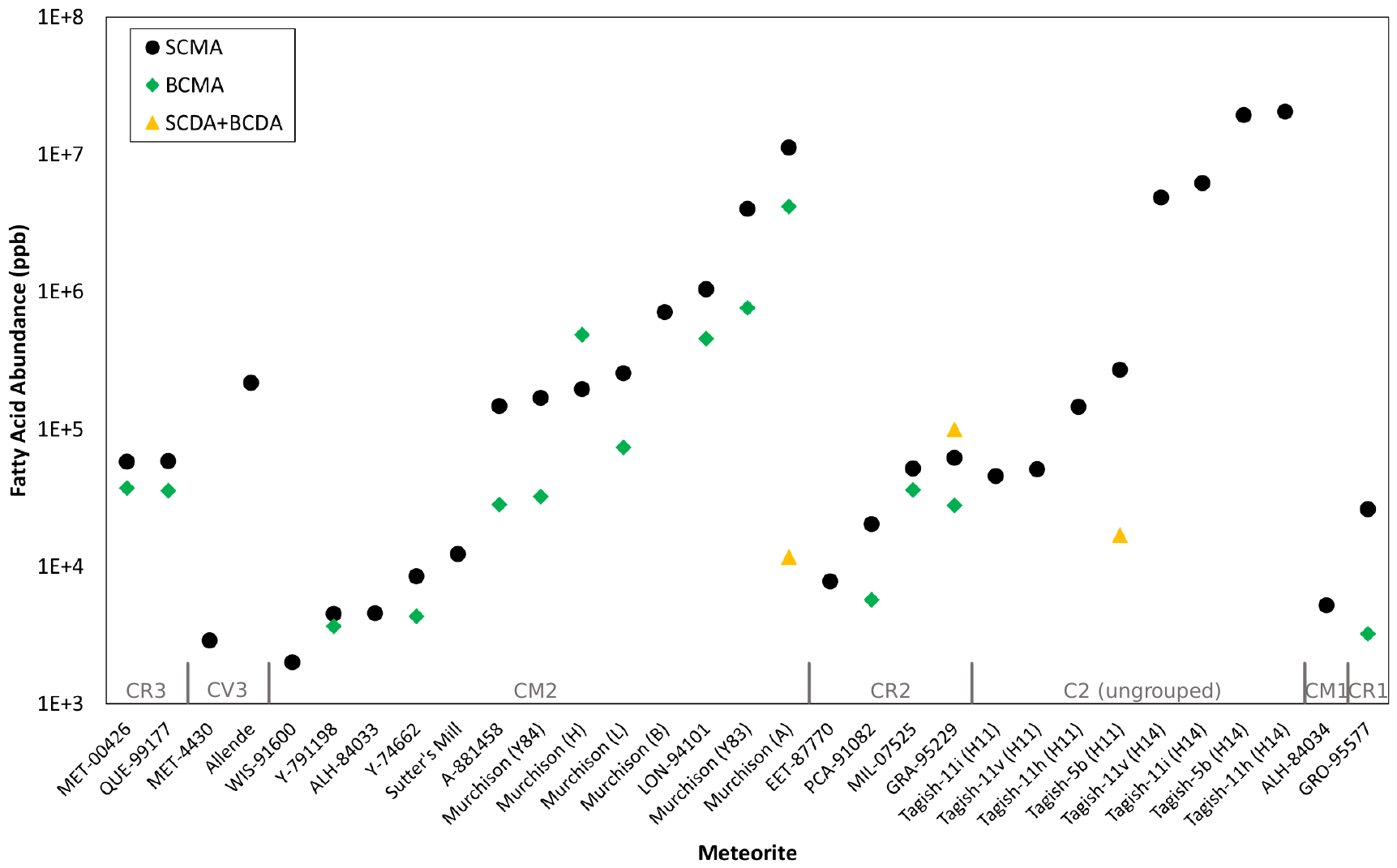}
	\caption{Total fatty acid abundances in all meteorite samples, expressed in parts-per-billion (ppb). Meteorites are first ordered by petrologic type (1, 2, or 3) and then by meteorite class (CV, CM, CR, or C (ungrouped)). SCMA: straight-chain monocarboxylic acids, BCMA: branched-chain monocarboxylic acids, SCDA: straight-chain dicarboxylic acids, BCDA: branched-chain dicarboxylic acids. \label{totalAb}}
\end{figure*}

In Figure~\ref{totalAb}, we display the sum of all SCMA, BCMA, and dicarboxylic acids in each meteorite, separated by petrologic type (1, 2, or 3) and meteorite type (CR, CV, CM, or C-ungrouped). In all but two cases, total SCMA abundances dominate over BCMA and dicarboxylic acid abundances. In one sample of the Murchison meteorite, BCMA abundances are greater than SCMA abundances. In the GRA-95229 meteorite, dicarboxylic acid abundances (SCDA + BCDA) were the most abundant fatty acids. Petrologic type 2 meteorites, on average, have the greatest abundances of fatty acids, dominated mostly by the CM2 and Tagish Lake (ungrouped) meteorites.

\subsection*{Amino Acids Compared with their Fatty Acid Decay Products}

\begin{figure*}
	\includegraphics[width=0.8\linewidth]{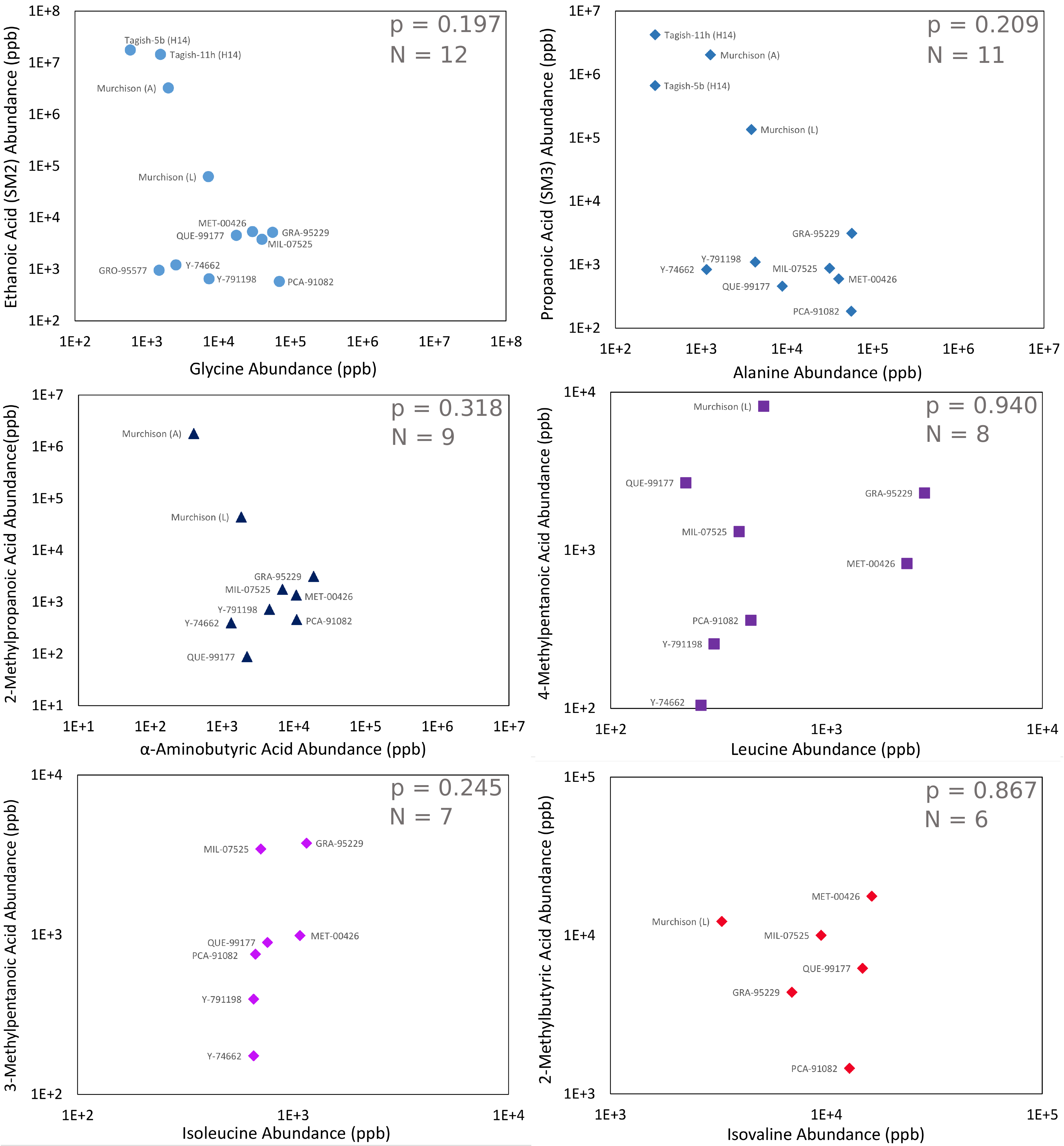}
	\caption{Correlation plots of meteoritic monocarboxylic acids to the corresponding meteoritic amino acids that differ only by a -NH$_2$ functional group. p-values correspond to the significance of anti-correlation in the top 4 frames, and correlation in the bottom 2 frames. N is the sample size. No correlations are found between amino acid abundances and their corresponding monocarboxylic acid decay product abundances with a significance of p $<$ 0.05. Meteorites are ordered by type (e.g. CR1, CR2, CR3, CM2, C2) and by decreasing abundance within each type. \label{Aminocomp}}
\end{figure*}

Amino acids decay into monocarboxylic acids through the loss of their amine functional group \citep{Reference420}. In Figure~\ref{Aminocomp}, we perform a Pearson correlation between amino acids and their monocarboxylic acid decay products in meteorites. We find no correlation or anti-correlation between the six displayed monocarboxylic acid/amino acid abundance pairs. This may suggest that the amino acid decay reaction (reaction no. 4 in Table~\ref{tbl-reactions}) is not the dominant mechanism to forming monocarboxylic acids within meteorite parent bodies. This result is in agreement with \citet{Reference447}, who explored the decay products of glycine and isovaline using ab initio molecular dynamics.

\end{document}